%% file: manuscript.tex
\documentclass[nofootinbib,10pt, twocolumn]{revtex4-1}
\usepackage[utf8]{inputenc}
\usepackage[T1]{fontenc}
\usepackage[paper=a4paper,left=25mm,right=25mm,top=25mm,bottom=20mm]{geometry}
\usepackage[english]{babel}
\usepackage{setspace}
\usepackage{physics}
\usepackage{mhchem}
\usepackage{mathtools}
\usepackage{bm}

\usepackage{graphicx}
\usepackage[stable]{footmisc}
\usepackage{xcolor}
\usepackage{soul}

\usepackage{hyperref} % hyperlinks for pdf navigation + coloring of references
\hypersetup{colorlinks=false, citebordercolor=green}

\date{\today}

\begin{document}
\title{Entanglement-assisted tunneling dynamics of impurities in a double well immersed in a bath of lattice trapped bosons}

\author{Friethjof Theel $^1$}
\author{Kevin Keiler $^1$}
\author{Simeon I. Mistakidis $^1$}
\author{Peter Schmelcher $^{1,2}$}
\affiliation{$^1$Center for Optical Quantum Technologies, University of Hamburg, Department of Physics, Luruper Chaussee 149, 22761 Hamburg, Germany}
\affiliation{$^2$The Hamburg Centre for Ultrafast Imaging, University of Hamburg, Luruper Chaussee 149, 22761 Hamburg, Germany}

\input{abstract}

\maketitle

\input{introduction.tex}

\input{setup_methodology.tex}

\input{correlated_tunnelling_dynamics.tex}

\input{effective_potential.tex}

\input{two_impurities.tex}

\input{conclusion.tex}

\input{appendix.tex}

\input{acknowledgements.tex}

\input{references.tex}
\end{document}

%% file: abstract.tex
\begin{abstract}
We unravel the correlated tunneling dynamics of an impurity trapped in a double well and interacting repulsively with a majority species of lattice trapped bosons. Upon quenching the tilt of the double well it is found that the quench-induced tunneling dynamics depends crucially on the interspecies interaction strength and the presence of entanglement inherent in the system. In particular, for weak couplings the impurity performs a rather irregular tunneling process in the double well. Increasing the interspecies coupling it is possible to control the response of the impurity which undergoes a delayed tunneling while the majority species effectively acts as a material barrier. For very strong interspecies interaction strengths the impurity exhibits a self-trapping behaviour. We showcase that a similar tunneling dynamics takes place for two weakly interacting impurities and identify its underlying transport mechanisms in terms of pair and single-particle tunneling processes.
\end{abstract}

%% file: introduction.tex
\section{Introduction}
Ultracold atoms offer a versatile platform for studying many-body effects in an extraordinarily controlled manner. Apart from varying the external confining potential and its dimensionality \cite{boshier,box_pot,diff_pot_spec}, it is also possible to tune the interaction strength between the atoms via Feshbach or confinement induced resonances \cite{confine,feshbach}.
This exquisite level of control over single component fermionic or bosonic ensembles can be extended to mixtures of ultracold atoms such as Bose-Bose \cite{pflanzer1,pflanzer2,phase_sep1,ofir,phase_sep2, phase_sep3,keiler3,pyzh,zollner,simos1,lia}, Fermi-Fermi\cite{ff1,ff2} and Bose-Fermi \cite{bf1,bf2} mixtures. In particular, one-dimensional systems exhibit intriguing phenomena since they allow for correlations to appear in the dilute regime  \cite{1d_1,1d_2,1d_eff_corr_1,1d_eff_corr_2}.\par
In this context, especially strongly particle imbalanced mixtures have attracted a lot of interest recently. In the extreme case such systems consist of a single impurity immersed in a majority species. These setups have been studied theoretically  \cite{blume_imp,cuc_imp,massignan_review,grusdt,volosniev,zinner_pol,garcia,jaksch_pol} and experimentally \cite{corn_pol_sim,naegerl_bloch,catani,fukuhura}, for a single impurity, serving as a simulator of polaron physics, as well as for many impurities \cite{zwerger_casimir,fleischhauer_ind,zinner_ind,jie_chen,keiler1,keiler2} and are indeed a subject of ongoing research. While the ground state properties of a single impurity in a bath are to a certain extent well understood, less focus has been placed on the transport properties and the emergent collisions of the impurity through the bath \cite{trans_imp,jaksch_clus,jaksch_trans,grusdt_simos}. Indeed, in these systems correlation effects, such as entanglement, are expected to be a crucial ingredient since the impurities form a few-body subsystem \cite{knoerzer}. Moreover, the underlying trapping potential plays an important role for the behaviour of the impurity species, which has been analyzed for homogeneous systems \cite{lampo,gamayun,lychkovskiy}, harmonic confinements \cite{akram,simosfimp,simosortho,simoseff,erdmann} as well as lattice potentials \cite{naegerl_bloch,kamar,siegl}. The majority of the above-mentioned investigations have been focusing on the case where both species are trapped in the same geometry. However, introducing different trapping potentials for each species is expected to alter significantly the observed dynamics. A setting of particular interest involves a bath of lattice trapped bosons which act as multiple material barriers for the tunneling dynamics of the impurity.\par
In the present work we explicitly focus on an impurity which is confined in a one-dimensional double well and interacts repulsively via contact interaction with a majority species of bosons trapped in a lattice. For single component bosons in a double well  the  analogue  of  the  well-known  superconducting Josephson junction can be established. The bosonic Josephson junction provides the testbed for many, also experimentally observed, intriguing  phenomena,  such  as  Josephson  oscillations,  macroscopic  quantum  self-trapping \cite{self1,self2,self3,self4,self5,self6} and correlated  pair  tunneling \cite{pair1,pair2,pair3}. Extensions of these phenomena to multicomponent setups have also been extensively studied, see for instance \cite{multijos1,multijos2,multijos3,multijos4}. Herein, we extend these investigations by exploring the dynamics of impurities in a double well immersed in a few-body bath of lattice trapped bosons. This gives rise to an effective potential for the impurities whose shape strongly depends on the interspecies interaction strength. Depending on the latter, one can realize tunneling scenarios which are beyond the well-known regimes of Josephson oscillation and quantum self-trapping and rely on the interspecies entanglement. This can be of particular interest for future applications in atomtronics {\cite{holland}}. \par
In our setup of a single impurity in a double well the dynamics is steered by the repulsive coupling to the majority species. Varying the interspecies interaction strength we unravel different dynamical response regimes of the impurity upon quenching the tilt of the double well. These regimes range from rather irregular tunneling in the double well for small interspecies interaction strengths to dynamical self-trapping in a single site for very strong couplings \cite{pflanzer1,pflanzer2,erdmann}. For intermediate coupling strengths we observe a strong impact of the density distribution of the majority species on the impurity's tunneling dynamics. The impurity initially collides with the material barrier imposed by the density of the majority species and then tunnels to the corresponding other site of the double well. This offers a controlled way of transporting the impurity within the double well. The entire tunneling process in the case of intermediate interspecies interaction strengths is accompanied by a strong entanglement between the subsystems revealing the complexity of this phenomenon. We remark that in the absence of entanglement this process does not take place. Additionally, in this case the self-trapping behaviour is altered. Surprisingly, we find that the dynamics of the impurity can be described in terms of Wannier states \cite{keiler1,keiler2} which are associated with the superposition of the effective time-averaged potential induced by the density of the majority species and the double well potential. This proves to be a valuable tool that captures the dynamics of the impurity adequately, even though a strong entanglement persists throughout the dynamics \cite{keiler3,grusdt_simos}. Due to the strong correlations appearing in our system it is necessary to utilize an approach which operates beyond lowest band and mean-field approximations, such as the Bose-Hubbard model or Gross-Pitaevskii approximation. Therefore we track the emergent non-equilibrium dynamics by employing the  Multi-Layer Multi-Configurational Time-Dependent  Hartree  Method  for  atomic  Mixtures (ML-MCTDHX) \cite{mlb1,mlb2,mlx} that enables us to capture all the important particle correlations. \par
Our work is structured as follows: In section \ref{sec:setup_methodology} we present the system under investigation and the employed computational methodology. In section \ref{sec:Tunneling Dynamics} we unravel the quench-induced tunneling dynamics of the impurity, revealing also the crucial role of the inter- and intraspecies correlations. Section \ref{sec:analysis} is dedicated to an in-depth characterization of the microscopic effects involved in the dynamical response of the impurity. We extend our results to the case of two weakly interacting impurities in section \ref{sec:two_impurities} and conclude with a summary of our findings and a discussion of future directions in section \ref{sec:Conclusion}.

%% file: setup_methodology.tex
\section{Setup and Multi-Configurational Approach}
\label{sec:setup_methodology}
\subsection{Setup and Hamiltonian}
Our setup consists of two different species of bosons $A$ and $B$, also referred to as the majority species and the impurity, respectively, which interact repulsively via a contact potential of strength $g_{AB}$. Each species is confined in a different one-dimensional optical potential at zero temperature. Experimentally this can be realized by preparing e.g. $^{87}{Rb}$ atoms in two different hyperfine states, i.e. $|F= 2,m_F=-2\rangle$ and  $|F= 1,m_F=-1\rangle$, thereby obtaining a  two-species  bosonic  mixture. Utilizing the  so-called  'tune-out' wavelength \cite{spec_sel_lat,mbl_hyperfine} it is possible to create species-dependent potentials, such that the two species experience  different optical potentials \cite{diff_pot_spec}. The majority A species, composed of bosons of mass $m_A$ and interacting repulsively via a contact interaction of strength $g_{AA}$, is trapped in a six-well lattice potential. The minority $B$ species on the other hand, consisting of $N_B$ impurities of mass $m_B$ interacting repulsively via a contact interaction of strength $g_{BB}$, resides in an initially tilted double well potential. The resulting many-body Hamiltonian of the system reads
\begin{align}\label{eq:Hamiltonian}
\mathcal{H} =&
\sum_{i=1}^{N_A} \Bigg( 
- \frac{\hbar^2}{2m_A} \frac{\mathrm{d}^2}{(\mathrm{d}x_i^{A})^2}
+ V_0\cos^2(k_0x_i^A) \Bigg.
\nonumber \\&
+ g_{AA} \Bigg.\sum_{i<j} \delta(x_i^A-x_j^A) \Bigg)
\nonumber \\&
+ \sum_{i=1}^{N_B} \Bigg( 
- \frac{\hbar^2}{2m_B}\frac{\mathrm{d}^2}{(\mathrm{d}x_i^{B})^2} \Bigg.
+ \frac{1}{2}m_B\omega_B^2(x_i^B)^2 
\nonumber \\&
+ \Bigg.\frac{h}{\sqrt{2\pi}w}\exp\!\left(-\frac{(x_i^B)^2}{2w^2}\right)
+ \alpha x_i^B\\ 
&+ g_{BB} \Bigg.\sum_{i<j} \delta(x_i^B-x_j^B) \Bigg)
\nonumber \\&
+ g_{AB} \sum_{i=1}^{N_A}\sum_{j=1}^{N_B}\delta(x_i^A-x_j^B).
\end{align}
Here, the lattice potential $V_A=V_0\cos^2(k_0x_i^A)$ of the majority species is characterized by its depth $V_0$ and wave vector $k_0=\pi/l$ where $l$ denotes the distance between two successive minima of the potential. The double well of the impurities $V_B=\frac{1}{2}m_B\omega_B^2(x_i^B)^2 + \frac{h}{\sqrt{2\pi}w} \exp\!\left(-\frac{(x_i^B)^2}{2w^2}\right)$ is constructed by the combination of a harmonic oscillator potential with frequency $\omega_B$ and a Gaussian potential characterized by a width $w$ and a height $h$. Additionally, we superimpose a linear tilting potential $V_{\textrm{tilt}}=\alpha x_i^B$ to the double well leading to an asymmetry between the two wells, whose degree can be controlled by the parameter $\alpha$. Assuming zero temperature we can model the inter and intraspecies interaction potential between the atoms via a bare delta potential with effective coupling strength
$g_{\sigma\sigma'} = \frac{2\hbar^2a^{\sigma\sigma'}_0}{M_{AB}a_\perp^2}
\left(1-\frac{\abs{\zeta(1/2)}a^{\sigma\sigma'}_0}{\sqrt{2}a_\perp} \right)^{-1} $
where $\sigma$ and $\sigma'$ refer to the corresponding species $A$ and $B$ \cite{confine}. Here, $M_{AB}=\frac{m_Am_B}{m_A+m_B}$ represents the reduced mass and $a_\perp=\sqrt{\frac{\hbar}{M_{AB}\omega_\perp}}$
the transversal length scale which is steered by the frequency of the transversal confinement $\omega_\perp$ perpendicular to the one-dimensional Bose gas.
Apart from varying $\omega_\perp$, it is possible to control the coupling strength $g_{\sigma\sigma'}$ through the free space, three-dimensional scattering length $a^{\sigma\sigma'}_0$ which can be tuned via Feshbach resonances in magnetic or optical fields \cite{feshbach,Inouye1998,Timmermans1999,Fedichev1996,Theis2004}.\par
Throughout this work we consider a fixed number of bosons for the majority species $N_A=8$ and set $m=m_A=m_B$. As mentioned above, our setup can be experimentally realized by considering two hyperfine states of \ce{^{87}Rb}. Note that we have also simulated the corresponding dynamics of a mass imbalanced system consisting e.g. of a \ce{^{87}Rb} bosonic ensemble and a \ce{^{133}Cs} impurity. For this latter case we confirmed that an overall similar phenomenology compared to the mass balanced case occurs but the emerging tunneling regimes to be presented below take place at smaller interspecies interaction strengths. The energy scales for the Hamiltonian in Eq. (\ref{eq:Hamiltonian}) are given in units of the recoil energy $E_r=\hbar^2k_0^2/(2m)$, whereas the length and time scales are expressed in units of $k_0^{-1}$ and $\omega_r^{-1}=\hbar E_r^{-1}$. For the lattice potential of the majority species we use a depth of $V_0/E_r=8$. The harmonic part of the double well potential of the impurities has a harmonic oscillator frequency of $\omega/\omega_r=0.1 \cdot \sqrt{2}$ and the barrier height and width are $h/E_rk_0^{-1}=2$ and $w/k^{-1}_0=1$, respectively. Furthermore, the intraspecies interaction strength among the bosons of the majority species is kept fixed to the value $g_{AA}/E_rk_0^{-1}=1$. Hard-wall boundary conditions are imposed at $x/k_0^{-1}=\pm3\pi$. \par
\begin{figure*}[t]
	\includegraphics[width=\textwidth]{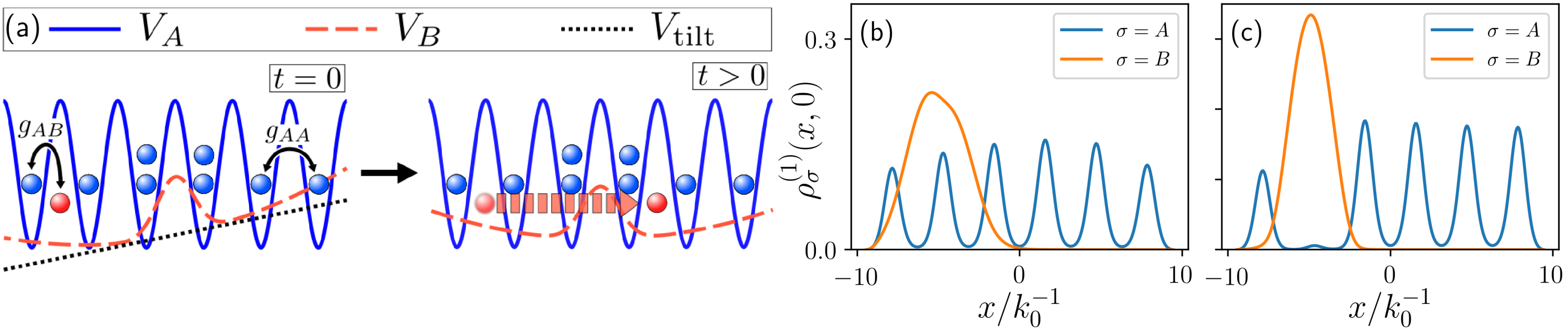}
	\caption{(a) Schematic representation of our setup before ($t=0$) and after the quench ($t>0$). The majority species (blue balls) resides in the lattice potential $V_A$. The impurity is embedded in a double well potential $V_B$ with an initially superimposed  tilt potential $V_{\textrm{tilt}}$ of strength $\alpha/E_rk_0^{-1} = 0.1$. The quench is performed by setting the tilting strength to zero, thereby quenching to a symmetric double well configuration. One-body density of the many-body ground state ($t=0$) of the species $\sigma$ for (b) $g_{AB}/E_r k^{-1}_0=0.2$ and (c) $g_{AB}/E_r k^{-1}_0=4.0$. We consider a minority species consisting of $N_B=1$ particle and a majority species composed of $N_A=8$ particles which interact repulsively with $g_{AA}/E_rk_0^{-1}=1.0$.}
	\label{fig:Setup}
\end{figure*}

%{\color{red}{[values in code, V0=4, h=1!, w=1, omegB=0.1, gA=0.5, $\alpha$=0.05]}}
In the following, we present the quench protocol which induces the tunneling dynamics. A sketch of the employed procedure is depicted in Figure \ref{fig:Setup}. First, we obtain the many-body ground state of our system, assuming the above mentioned parameters. Here, the tilting strength of the double well is set to $\alpha/E_rk_0^{-1}=0.1$ (the effect of a smaller tilting strength is analyzed in the Appendix), such that the impurities localize in the left well of the asymmetric double well potential. To trigger the tunneling dynamics of the impurities the system is quenched to a geometry, constituting a symmetric double well, i.e. the tilting strength is set to $\alpha=0$. Varying the interspecies interaction strength $g_{AB}$, we explore the dependence of the system dynamics on $g_{AB}$.
\subsection{Approach to the correlated many-body dynamics}
To unravel the dynamics of the system we employ ML-MCTDHX \cite{mlb1,mlb2,mlx}. As explicated below, this \textit{ab initio} method gains its efficiency from the time-dependent and with the system co-moving basis set. In the first step, the total many-body wave function $\ket{\Psi_{\textrm{MB}}(t)}$ is expanded with respect to $M$ different species functions $\ket{\Psi^{\sigma}(t)}$ for each of the species $\sigma$ and expressed according to the following Schmidt decomposition \cite{schmidt_dec} 
\begin{equation}
\ket{\Psi_{\textrm{MB}}(t)} = \sum_{i=1}^{M} \sqrt{\lambda_{i}(t)} \ket{\Psi_{i}^A(t)}\otimes \ket{\Psi_{i}^B(t)}.
\label{eq:schmidt}
\end{equation}
Here, the Schmidt coefficients $\sqrt{\lambda_{i}(t)}$, in decreasing order, provide information about the degree of population of the $i$-th species function and thereby signify the degree of entanglement between the two species. In the case that only one Schmidt coefficient is non-zero, the species $A$ and $B$ are not entangled with each other and the system can be described by a species mean-field ansatz ($M=1$). However, in general it is necessary to provide several species functions for the expansion of the total many-body wave function, since entanglement might prove crucial for the adequate description of the systems dynamics.\par
Furthermore, the species wave functions $\ket{\Psi^{\sigma}(t)}$ describing an ensemble of $N_\sigma$ bosons are expanded in a set of permanents, namely
\begin{align}
	\ket{\Psi_{i}^\sigma(t)} = \sum_{\vec{n}^\sigma|N_\sigma} C_{\sigma\vec{n}}(t)
	|\vec{n}^\sigma(t)\rangle.
\label{eq:ns}
\end{align}
Such an expansion allows us to take intraspecies correlations of the $\sigma$-species into account. Moreover, in this  expression the vector $\vec{n}^\sigma=(n^{\sigma}_1,n^{\sigma}_2,...)$ describes the occupations of the time-dependent single-particle functions (SPF) of the species $\sigma$, which are further expanded in terms of a time-independent discrete variable representation \cite{Light1985}. The notation $\vec{n}^\sigma|N_\sigma$ indicates that for each $n_i^\sigma$ the particle number conservation condition $\sum_{i}n^{\sigma}_i=N_\sigma$ has to be fulfilled. For the time propagation of the many-body wave function we employ the Dirac-Frenkel variational principle
$ \bra{\delta\Psi_\textrm{MB}} (\textrm{i}\partial_t - \mathcal{H} )\ket{\Psi_\textrm{MB}} $ \cite{var1,var2,var3} with the variation $\delta\Psi_\textrm{MB}$ and obtain the corresponding equations of motion \cite{mlx,pruning}.\par
In conclusion, the ML-MCTDHX method takes all inter- and intraspecies correlations into account and gives us access to the complete many-body wave function. In contrast to standard approaches, where the wave function for solving the time-dependent Schrödinger equation is built upon time-independent Fock states with time-dependent coefficients, the ML-MCTDHX method takes a co-moving time-dependent basis into account, where the Fock states, spanned by the SPFs, as well as the coefficients are time-dependent. This concept of a time-dependent basis reduces not only the required number of basis states and, hence, improves the computational effort, but it also provides at the same time an accurate description of the system's many-body state. We note here that in order to ensure the convergence of our many-body simulations, to be presented below, we have employed $M=6$ ($M=10$) species and $d_A=6$, $d_B=6$ ($d_A=6$, $d_B=6$) single-particle functions respectively for the case of a single (two) impurity atom(s). In this context we define the orbital configuration $C=(M,d_A,d_B)$ which determines the size of the truncated Hilbert space. 

%% file: correlated_tunnelling_dynamics.tex
\section{Correlated Tunneling Dynamics of a Single Impurity}
\label{sec:Tunneling Dynamics}
\begin{figure*}[t]
	\includegraphics[scale=0.55]{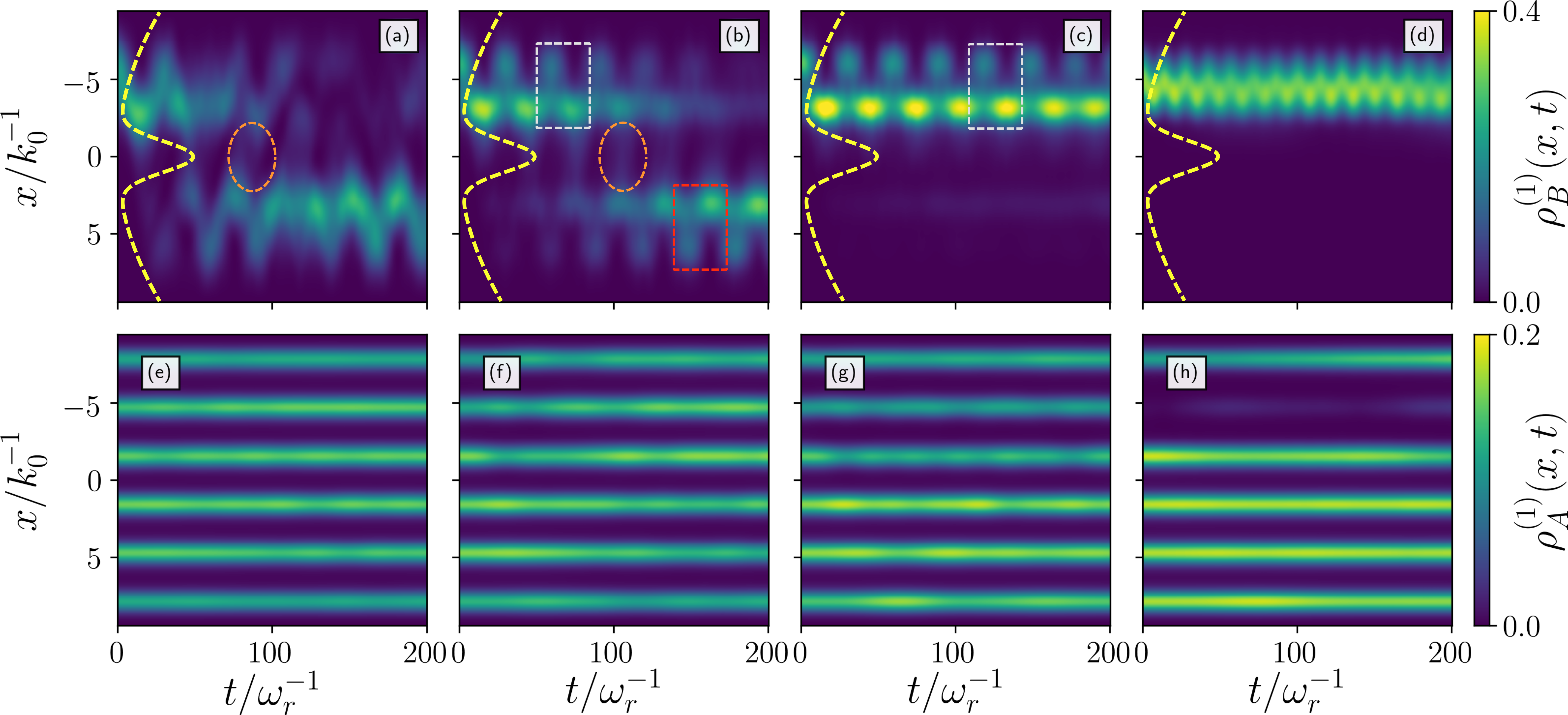}
	\caption{Temporal evolution of the one-body densities for (a)-(d) the impurity species B and (e)-(h) the majority species A upon quenching the tilting strength from $\alpha/E_R k^{-1}_0=0.1$ to $\alpha/E_R k^{-1}_0=0$. Each column corresponds to a different interspecies interaction strength $g_{AB}$, from left to right for $g_{AB}/E_R k^{-1}_0=0.2,1.0,2.0,4.0$. The dashed yellow line represents the double well potential for the impurity species.  We consider a minority species consisting of $N_B=1$ particle and a majority species of $N_A=8$ particles which interact repulsively with $g_{AA}/E_rk_0^{-1}=1.0$.}
	\label{fig:MB_densities}
\end{figure*}
In the following we consider a mass-balanced bosonic mixture described by the Hamiltonian of equation \ref{eq:Hamiltonian} where the majority species consists of $N_A=8$ and the impurity species of $N_B=1$ particles. We initially prepare our system in its ground state with a tilting strength of $\alpha/E_R k^{-1}_0=0.1$ for different interspecies interaction strengths $g_{AB}$. Due to the initial tilt the impurity is found to be well localized in a single site of the double well potential. Moreover, we find that the impurity species exhibits a rather large spatial overlap with the majority species for small $g_{AB}$ [see Figure \ref{fig:Setup} (b)] which of course reduces with increasing repulsive $g_{AB}$ [see Figure \ref{fig:Setup} (c)]. In particular, for small $g_{AB}$ the majority species occupies all sites of the lattice potential, such that the impurity strongly overlaps with it [cf. Figure \ref{fig:Setup} (b)]. However, for strong repulsive interactions the majority species depopulates the well of the lattice potential in which the impurity tends to localize, resulting in a weak spatial overlap of the two species [cf Figure \ref{fig:Setup} (c)]. Upon quenching the tilting strength to $\alpha/E_R k^{-1}_0=0$ towards a symmetric double well we initiate the tunneling dynamics, thus favoring the tunneling of the impurity to the right well as the corresponding energy offset between the two wells vanishes, see also Fig. 1 (a). As a consequence the impurity becomes mobile, thereby colliding with the lattice trapped majority species which in general acts as a material barrier for the impurity dynamics. Varying the interspecies interaction strength we find four different regimes for the dynamical response of the impurity (see below).\par
As a first step, we quantify these regimes by monitoring the time evolution of the one-body density $\rho^{(1)}_{\sigma}(x,t)=\langle\Psi_\text{MB}(t)| \hat{\Psi}_{\sigma}^{\dagger}(x)\hat{\Psi}_{\sigma}(x)|\Psi_\text{MB}(t) \rangle$  of the corresponding subsystems $\sigma$. The spectral decomposition of the $\sigma$-species one-body density is given by
\begin{equation}
\rho_\sigma^{(1)}(x,t) = \sum_j n_{\sigma j}(t) \Phi^{*}_{\sigma j}(x,t)\Phi_{\sigma j}(x,t),
\label{eq:natural_populations}
\end{equation}
where $n_{\sigma j}(t)$ are the so-called natural populations and $\Phi_{\sigma j}(x,t)$ the corresponding natural orbitals.
The dynamics of $\rho_\sigma^{(1)}(x,t)$ is presented in Figure \ref{fig:MB_densities}, for different interspecies interaction strengths $g_{AB}$. As it can be seen $\rho_\sigma^{(1)}(x,t)$ exhibits four distinct dynamical response regimes. For small interspecies interaction strengths, in our case $g_{AB}/E_R k^{-1}_0=0.2$, the impurity undergoes a rather complex tunneling dynamics to the other site of the double well [Figure \ref{fig:MB_densities} (a)]. This is a single-particle effect caused by the strong initial tilt and is therefore also present for $g_{AB}=0$. For short evolution times, i.e. $0<t/\omega^{-1}_r<50$, the impurity performs oscillations in the initial well and then tunnels [see ellipse in Figure \ref{fig:MB_densities} (a)] to the other well.  Here, the oscillations within each of the two wells, which still persist even for $g_{AB}=0$ (not shown here), are caused by the rather strong initial tilt of the double well and are not present for smaller tilts \footnote{We note that this tunneling behaviour differs from that of a single particle in a double well potential in the case of smaller tilts, yielding a single frequency Rabi tunneling.} [see Figure \ref{fig:MB_densities_small_tilt} (a)]. In this sense, the majority species barely affects the tunneling dynamics of the impurity and exhibits weak amplitude modulations from its initial profile due to the finite $g_{AB}$ [Figure \ref{fig:MB_densities} (e)].\par
However, for larger coupling strengths the impurity is strongly influenced by the density distribution of the majority species, e.g. see Figure \ref{fig:MB_densities} (b),(f). The majority species distributes over the lattice such that $\rho_A^{(1)}(x,t)$ is accumulated close to the minima of the lattice potential. Due to the repulsive interspecies interaction the impurity has to overcome  on top of the double well barrier these additional material barriers imposed by the accumulation of the density of the majority species. This leads to an oscillation of the impurity through the neighboring density maximum of the A species [see white rectangle in Figure \ref{fig:MB_densities} (b)]. This tunneling through the material barrier imposed by the majority species we  will refer to as material barrier tunneling in the following. Throughout this enduring oscillation process the impurity performs a transport \cite{overbarrier1,overbarrier2} to the other site of the double well [see ellipse in Figure \ref{fig:MB_densities} (b)] where it again encounters a material barrier of species A and pursues the initial material barrier tunneling behavior [see red rectangle in Figure \ref{fig:MB_densities} (b)]. Compared to the weakly interacting regime [Figure \ref{fig:MB_densities} (a)], in the intermediate regime the transport of the impurity to the other site of the double well takes place in a very controlled and systematic manner. Moreover, it is even possible to prolong the initial material barrier tunneling process by further increasing the interspecies interaction strength [Figure \ref{fig:MB_densities} (c)]. In this case, the impurity undergoes a weak amplitude tunneling to the other site of the double well [cf. Figure \ref{fig:Wannier_states_occupation} (c)], at least within the considered evolution time. In the limit of very large $g_{AB}$ the impurity is trapped in the initial site of the double well due to the strong interspecies repulsion [cf. Figure \ref{fig:Setup} (c) and Figure \ref{fig:MB_densities} (d)] and as a result we enter the self-trapping regime. We remark that this self-trapping behavior is caused by the presence of the majority species, in sharp contrast to the well-known case of interacting bosons in a double well. Note also that the impurity undergoes dipole-like oscillations within the left site of the double well. Also, we have checked that this self-trapping behavior [cf. Figure \ref{fig:MB_densities} (d)] of the impurity persists up to $t/\omega^{-1}_r=400$ evolution times (not shown here).\par
Considering the behavior of the majority species A, we observe the development of excitations of $\rho_A^{(1)}(x,t)$ as a back-action of the tunneling process of the impurity \cite{keiler3}. In particular, $\rho_A^{(1)}(x,t)$ is transferred through the lattice [Figure \ref{fig:MB_densities} (f), (g)]. Predominantly, this is the case for the inner four wells. This behaviour of the majority species is caused by the repulsive interspecies interaction which leads in the course of the impurity tunneling to a shift of the density of species A, thereby reducing the overlap between the species. In the extreme case [cf. $g_{AB}/E_R k^{-1}_0=4.0$] where the impurity remains localized in one site of the double well, the majority species redistributes such that a density hole is formed in one lattice site [cf. Figure \ref{fig:MB_densities} (h)], in order to avoid the impurity. Here, the overall density of the majority species barely changes in time due to the absence of the impurity's tunneling. \par
In order to quantify the dynamical response of the system even further it is convenient to analyze how strongly the time-dependent many-body wave function deviates from the initial state $|\Psi_0\rangle$ at  $t=0$ in the course of time. This is well captured by the fidelity $F(t)=|\langle\Psi_\text{MB}(t)|\Psi_0\rangle|^{2}$ which is defined as the overlap between the time-dependent and the initial wave function. 
\begin{figure}[t]
	\includegraphics[scale=0.55]{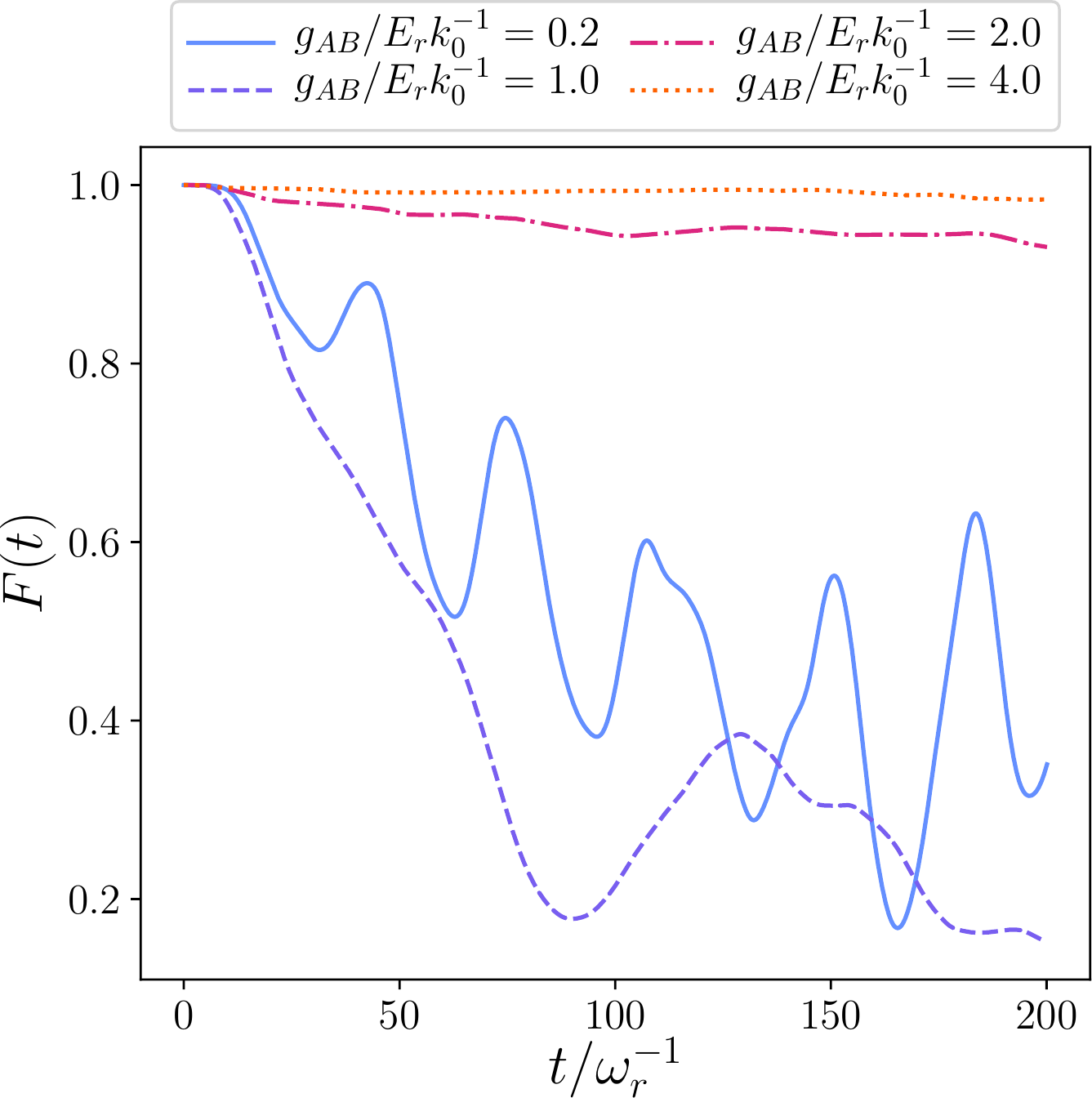}
	\caption{Temporal evolution of the fidelity $F(t)$ for different interspecies interaction strengths $g_{AB}$ upon quenching the tilting strength to $\alpha/E_R k^{-1}_0=0$. We consider a minority species consisting of $N_B=1$ particle and a majority species consisting of $N_A=8$ particles which interact repulsively with $g_{AA}/E_rk_0^{-1}=1.0$.}
	\label{fig:fidelity}
\end{figure}
\begin{figure*}[t]
	\includegraphics[width=.75\textwidth]{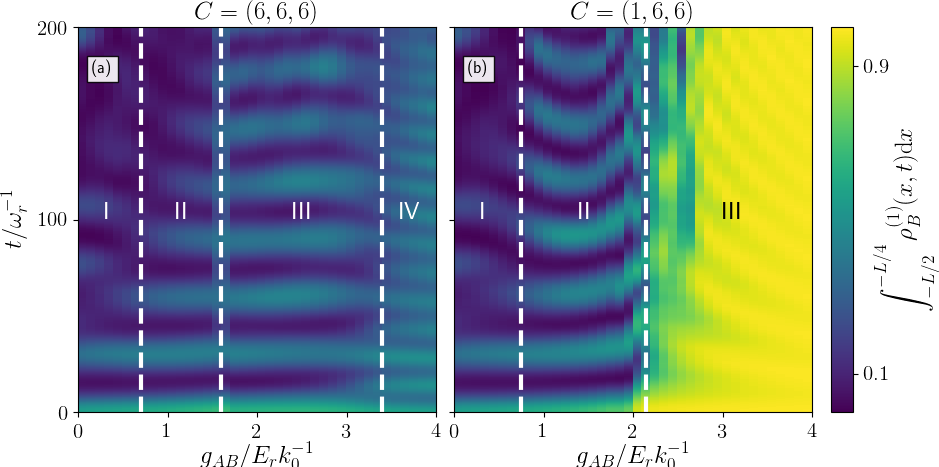}
	\caption{Temporal evolution of the integrated one-body density of the impurity $\int_{-L/2}^{-L/4}\rho_B^{(1)}(x,t) dx$ upon variation of the interspecies interaction strength $g_{AB}$ for (a) the full many-body approach and (b) the species mean-field approach. The regimes I-IV in panel (a) correspond to the one-body densities in Figure {\ref{fig:MB_densities}} (a)-(d). Regimes I,III in panel (b) correspond to the one-body densities in Figure {\ref{fig:MF_densities}} (a),(d), respectively, whereas regime II in panel (b) relates to Figure {\ref{fig:MF_densities}} (c) and (d). We consider a minority species consisting of $N_B=1$ particle and a majority species of $N_A=8$ particles which interact repulsively with $g_{AA}/E_rk_0^{-1}=1.0$.}
	\label{fig:integrated_density}
\end{figure*}
Figure \ref{fig:fidelity} shows the fidelity $F(t)$ for various interspecies interaction strengths corresponding to the four different tunneling regimes identified in the time evolution of the one-body densities in Figure \ref{fig:MB_densities}. We clearly observe that the behavior of the fidelity is qualitatively driven by the one-body density distribution of the impurity over time. For the cases in which the impurity tunnels to the other site of the double well [Figure \ref{fig:MB_densities} (a),(b)], the fidelity deviates significantly from unity, i.e. $|\Psi_\text{MB}(t)\rangle$ deviates from the ground state $|\Psi_0\rangle$. However, in the regimes where the tunneling of the impurity is suppressed the fidelity remains close to unity, e.g. see $F(t)$ for $g_{AB}=2.0,4.0$. In this sense, the fidelity evolution provides an indicator of the tunneling process of the impurity and serves as a first characterization for the tunneling [Figure \ref{fig:MB_densities} (a) and (b)] and self-trapping regimes [Figure \ref{fig:MB_densities} (c) and (d)]. Nevertheless, using solely the fidelity it is not possible to distinguish between the different tunneling mechanisms. In order to discern between the above-mentioned four possible regimes of the impurity's dynamical response it is useful to consider the integrated one-body density of the impurity $\int_{-L/2}^{-L/4}\rho_B^{(1)}(x,t) dx$, where $L$ is the size of the system. This quantity provides the probability of finding the impurity in one half of the initially populated well of the double well. Indeed, the integrated density allows to distinguish between the emergent tunneling dynamics since it incorporates the effect of the material barrier. In Figure {\ref{fig:integrated_density}} (a) we show the temporal evolution of this quantity for the correlated many-body approach \footnote{The many-body approach refers to our treatment within ML-MCTDHX in contrast to mean-field approaches.} for different $g_{AB}$. We clearly observe four distinct regimes for the response of the impurity which correspond to the one-body densities in Figure  {\ref{fig:MB_densities}} (a)-(d). Indeed, in regime I an irregular oscillatory pattern of the integrated density is found. Regime II exhibits regular oscillations whose intensity decays in time. This corresponds to the material barrier tunneling with a final transfer of the impurity to the other site of the double well [see {\ref{fig:MB_densities}} (b)]. In regime III the oscillations of the integrated density remain stable in time which is due to the material barrier tunneling of the impurity in the initially populated well without a transfer to the other site. Finally, in regime IV we find a higher-frequency  oscillatory behavior with a finite amplitude throughout the evolution. This behavior corresponds to the self-trapping regime [see Figure {\ref{fig:MB_densities}} (d)]. Regarding the dependence of the tunneling behavior on the different system parameters see Appendix {\ref{appB}}. \par
\begin{figure}[t]
	\includegraphics[scale=0.55]{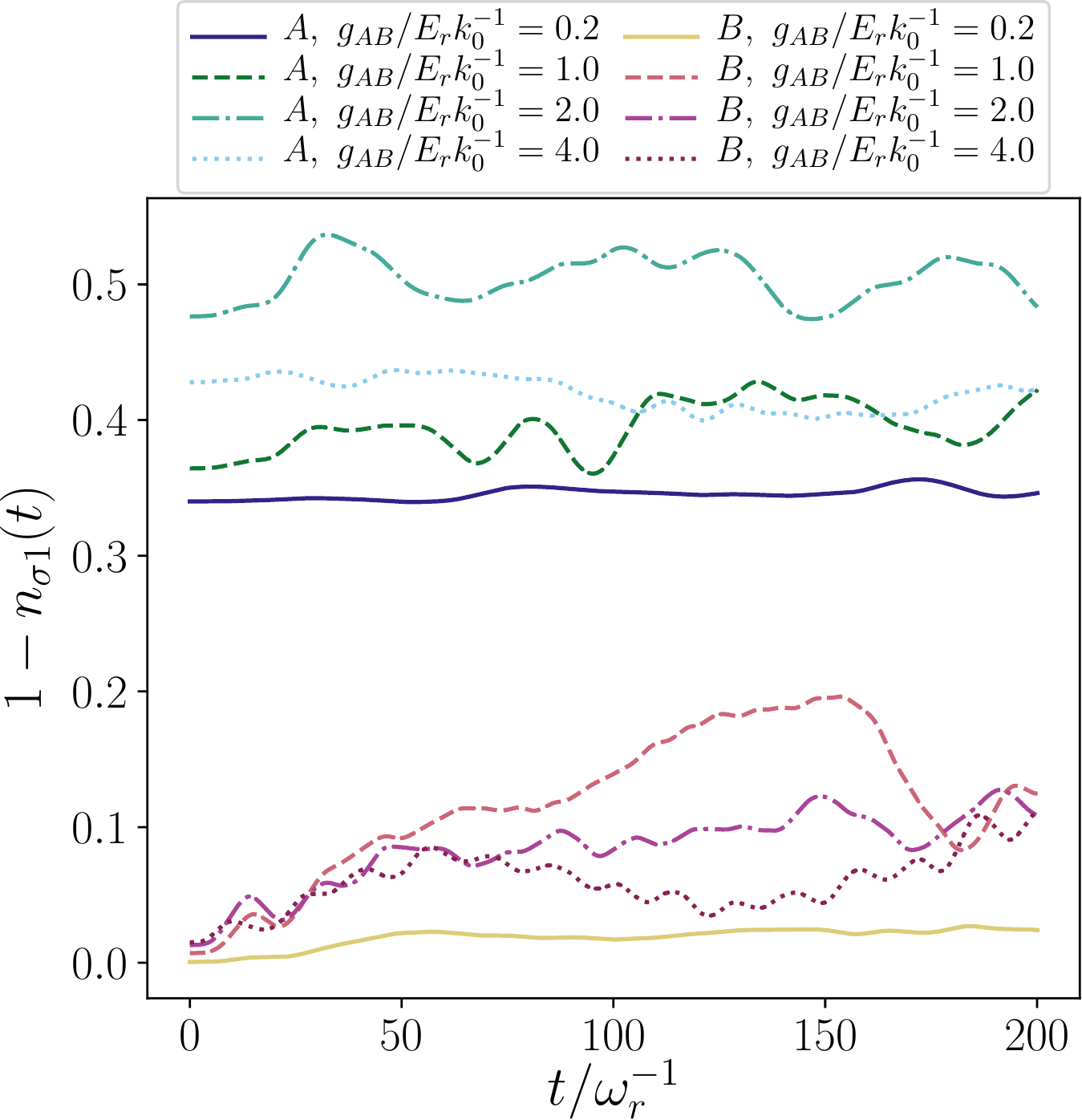}
	\caption{Temporal evolution of the depletion $1-n_{\sigma1}(t)$ of the most populated natural orbital of the A and the B species for different interspecies interaction strengths $g_{AB}$ upon quenching the tilting strength to $\alpha/E_R k^{-1}_0=0$. Note that the natural populations of the B species coincide with the Schmidt coefficients since $N_B=1$, thereby describing also the degree of entanglement between the subsystems. We consider a minority species consisting of $N_B=1$ particle and a majority species composed of $N_A=8$ particles which interact repulsively with $g_{AA}/E_rk_0^{-1}=1.0$.}
	\label{fig:depletion}
\end{figure}

However, so far we did not get insight into the degree of the system's correlation throughout the dynamics.  
To unravel the degree of correlations which accompanies the tunneling dynamics of the impurity we distinguish between inter- and intraspecies correlations. The former are described by the Schmidt coefficients (Eq. \ref{eq:schmidt}), which provide a measure for the degree of entanglement between the subsystems, whereas the latter can be inferred from the natural populations (cf. Eq. \ref{eq:natural_populations}). Since the B species consists of a single particle, the natural populations of the B species coincide with the Schmidt coefficients. Therefore, in the following we invoke the deviation $1-n_{B1}(t)$ as a measure of entanglement between the subsystems. Accordingly, $1-n_{A1}(t)$ indicates the degree of intraspecies correlations of the majority species.  The temporal evolution of the depletion $1-n_{\sigma1}(t)$ of the most populated natural orbital of the A and the B species is illustrated in  Figure \ref{fig:depletion} for different $g_{AB}$ upon quenching the tilting strength.
We observe that for small interspecies interaction strengths, i.e. $g_{AB}=0.2$, the subsystems are mainly disentangled throughout the dynamics, since $1-n_{B1}(t)\approx0$. Increasing the interspecies interaction strength to $g_{AB}/E_R k^{-1}_0=1.0$ the subsystems become strongly entangled in the course of time, i.e. $1-n_{B1}(t)>0$. This can be associated with tunneling of the impurity to the other site of the double well and the involved increasing interspecies interaction between the subsystems. Naturally, the motion of the impurity through the majority species has an impact on the natural populations of the A species which is connected to the intrinsic tunneling processes of the A species in the lattice potential [cf. Figure \ref{fig:MB_densities} (e)-(f)]. Indeed, $1-n_{A1}>0$ independently of $g_{AB}$ and it is maximized in the above-described third tunneling region. Interestingly, the rather strong degree of entanglement remains in the self-trapping regime for $g_{AB}/E_R k^{-1}_0=4.0$, even though the impurity barely overlaps with the majority species. \par
In order to emphasize the importance of the entanglement between the subsystems for the tunneling behavior of the impurity, we additionally perform calculations assuming only a single product state $\ket{\Psi_{\textrm{MB}}} = \ket{\Psi_{\textrm{A}}} \otimes \ket{\Psi_{\textrm{B}}}$ in Eq. (\ref{eq:schmidt}), thereby neglecting all interspecies correlations.
\begin{figure*}[t]
	\includegraphics[scale=0.55]{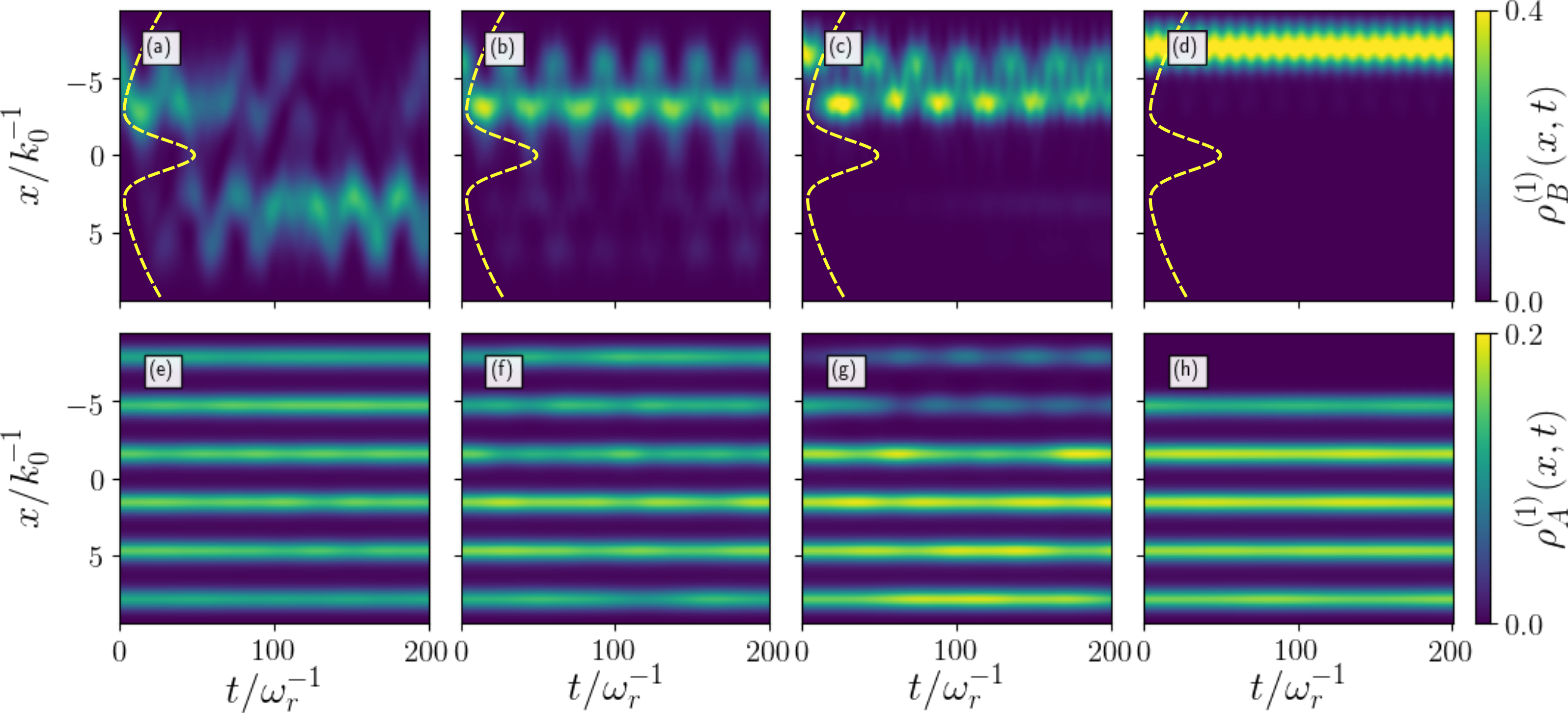}
	\caption{Temporal evolution of the one-body densities for (a)-(d) the impurity species B and (e)-(h) the majority species A upon quenching the tilting strength from $\alpha/E_R k^{-1}_0=0.1$ to $\alpha/E_R k^{-1}_0=0$, assuming a species mean-field ansatz. Each column corresponds to a different interspecies interaction strength $g_{AB}$, ranging from left to right with $g_{AB}/E_R k^{-1}_0=0.2,1.0,2.0,4.0$.  The dashed yellow line represents the double well potential for the impurity species. We consider a minority species consisting of $N_B=1$ particle and a majority species consisting of $N_A=8$ particles which interact repulsively with $g_{AA}/E_rk_0^{-1}=1.0$.}
	\label{fig:MF_densities}
\end{figure*}
The dynamics of the $\sigma$-species one-body densities employing a species mean-field ansatz, i.e. assuming a single product state between the species but still including intraspecies correlations, are shown in Figure \ref{fig:MF_densities}. For  $g_{AB}/E_R k^{-1}_0=0.2$ we find no visible differences between the full many-body approach and the species mean-field calculations. This is an expected result, as the degree of entanglement is rather small for these interactions (cf. Figure \ref{fig:depletion}). However, as soon as entanglement becomes important, we find strong deviations in the corresponding one-body densities. In particular, for $g_{AB}/E_R k^{-1}_0=1.0$ in the species mean-field case [cf. Figure \ref{fig:MF_densities} (b)] we do not observe the previously predicted tunneling to the other site of the double well [cf. Figure \ref{fig:MB_densities} (b)]. Furthermore, the one-body density of the impurity for $g_{AB}/E_R k^{-1}_0=2.0$ exhibits additional oscillation frequencies in the species-mean field scenario [cf. Figure \ref{fig:MF_densities} (c)] compared to the full many-body case [\ref{fig:MB_densities} (c)]. In the self-trapping regime, $g_{AB}/E_R k^{-1}_0=4.0$, the species mean-field calculations seem to capture the dynamics quite well at first glance. However, on a closer inspection of the one-body density it turns out that the spatial position of the impurity differs compared to the complete many-body approach, while the temporal oscillations of the density are also of different amplitude and frequency, see Figures \ref{fig:MB_densities} (d) and \ref{fig:MF_densities} (d). This general difference in the tunneling dynamics is well captured by the integrated density shown in Figure {\ref{fig:integrated_density}}. Indeed, the species mean-field ansatz is not able to recover regime II in Figure {\ref{fig:integrated_density}} (a), while the self-trapping regime III in Figure {\ref{fig:integrated_density}} (b) is strongly altered compared to that one in Figure {\ref{fig:integrated_density}} (a) corresponding to regime IV in the many-body treatment. Even regime III in Figure {\ref{fig:integrated_density}} (a) is quantitatively changed when using the species mean-field ansatz [cf. Figure {\ref{fig:integrated_density}} (b) regime II].  In this sense, entanglement between the impurity and the majority species plays a crucial role, in order to describe the dynamics correctly.

%% file: effective_potential.tex
\section{Characterization of the impurity dynamics}
\label{sec:analysis}
To analyze the tunneling behavior of the impurity and the accompanying correlations due to the presence of the majority species (cf. Figure \ref{fig:MB_densities}) we next develop an effective potential model for the impurity. This effective potential is obtained by superimposing the time-averaged density of the majority species to the external double well potential. To adequately describe the dynamical response of the majority species we employ the associated Wannier functions. In particular, we project the complete many-body wave function obtained via ML-MCTDHX onto these Wannier functions in order to analyze the behavior of the impurity in a fixed basis set.

\subsection{Construction of the effective potential}
\label{sec:Effective potential}
Initially, we prepare our system such that it is given by the ground state of the Hamiltonian (eq. \ref{eq:Hamiltonian}) with an underlying asymmetric double well. With respect to the quenched Hamiltonian ($t>0$) our system and in particular the impurity is energetically excited due to the tilting. This enables the impurity to tunnel through the potential barrier of the double well into the right well. However, as mentioned in section \ref{sec:Tunneling Dynamics}, for specific interspecies interaction strengths this residual energy appears to be not large enough to overcome the potential barrier. Therefore, the impurity $B$ rather performs a tunneling in the initial site of the double well through the material barrier imposed by the one-body density of the majority species [cf. Figure \ref{fig:MB_densities} (c)]. \par 
In the following, we aim at understanding this tunneling behavior using an effective potential for the impurity. We remark that this effective potential serves only as a tool for an analysis of the underlying tunneling processes. Integrating out the majority species we arrive at the following effective potential for the impurity
\begin{equation}
V_{\mathrm{eff}}^B(x^B,t) = N_A g_{AB} \rho_A^{(1)}(x^B,t)+V_B(x^B).
\label{eq:effective_potential}
\end{equation}
This effective potential is composed by the double well potential $V_B$ and the one-body density of the majority species $\rho_A^{(1)}$ being weighted by the number of particles $N_A$ and the interspecies interaction strength $g_{AB}$. Note that $\rho_A^{(1)}(x^B,t)$ is calculated within the correlated many-body approach and thereby includes all necessary correlations. $\rho_A^{(1)}(x^B,t)$ cannot be recovered within a mean-field approach. To proceed, we average this effective potential over time and obtain a Time-Averaged Effective Potential (TAEP)
\begin{equation}
\overline{V}^B_{\mathrm{eff}}(x^B)=\frac{1}{T}\int_0^T V_{\mathrm{eff}}^B(x^B,t)\textrm{d}t,
\label{eq:TAEP}
\end{equation}
where $T$ denotes the total propagation time. We can justify this time-average by the small one-body density deformations of the majority species over time. Furthermore, we remark that Eq. \ref{eq:effective_potential} is a species mean-field effective potential and, therefore, only assumes a single product state. Even though we have seen in Figure \ref{fig:depletion} that the entanglement between the subsystems plays a crucial role this ansatz turns out to be a powerful tool to analyze the basic aspects of the tunneling behavior of the impurity and gives an intuitive understanding \cite{grusdt_simos,keiler3,hannes_simos}.
\begin{figure*}[t]
	\centering
	\includegraphics[width=\textwidth]{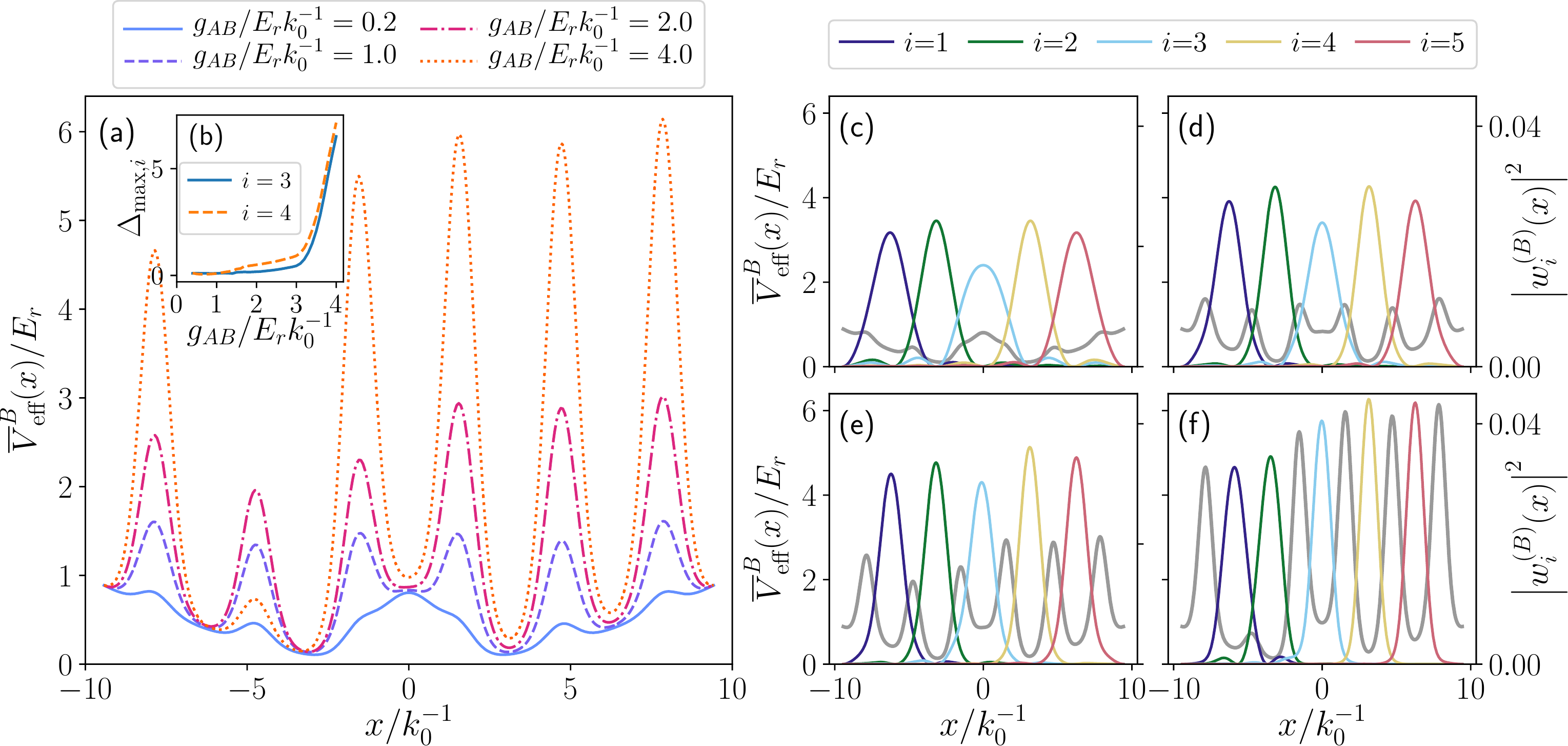}
	\caption{(a) Time-Averaged Effective Potential (TAEP) [Eq. \ref{eq:TAEP}] for the four different interspecies interaction strengths $g_{AB}$ corresponding to the four tunneling regimes of the impurity. (b) The relative difference of the height between the second maxima and the third and fourth maxima of the TAEP $\Delta_{\textrm{max},i}$ (see main text for definition) depending on $g_{AB}$.
	(c) The TAEP (gray lines) for $g_{AB}/E_rk_0^{-1}=0.2$ together with the corresponding Wannier functions which are calculated from the effective Hamiltonian constructed by the TAEP. (d)-(f) The same as in (c), but for different $g_{AB}$'s, viz. $g_{AB}/E_rk_0^{-1}=1.0$, $g_{AB}/E_rk_0^{-1}=2.0$ and $g_{AB}/E_rk_0^{-1}=4.0$, respectively.}
	\label{fig:effective_potential}
\end{figure*}
The time-averaged effective potentials are depicted in Figure \ref{fig:effective_potential} (a) for the four interspecies interaction strengths $g_{AB}/E_rk_0^{-1}=0.2,1.0,2.0,4.0$ corresponding to the four tunneling regimes already discussed in section \ref{sec:Tunneling Dynamics}. For weak interspecies interaction strengths, e.g. $g_{AB}/E_rk_0^{-1}=0.2$, the TAEP resembles the shape of the double well potential $V_B$, and only small deviations are visible raised from the superimposed one-body density of the majority species. Moreover, the height of the central potential barrier of the TAEP ($\approx 0.89 E_r$) is approximately the same as for the double well $V_B$ with a value of $h/(\sqrt{2\pi}w) \approx 0.80$ at $x^B=0$. Therefore, one can assume that the tunneling dynamics of the impurity in the TAEP would differ only marginally compared to the behavior of a single particle confined in the double well $V_B$, i.e. for $g_{AB}=0$. Since in the TAEP the one-body density $\rho_A^{(1)}$ of the majority species is weighted by the interspecies interaction strength $g_{AB}$, the spatial distribution of $\rho_A^{(1)}$ becomes more pronounced for increasing $g_{AB}$. Consequently, for a larger $g_{AB}$ we observe the appearance of six maxima on top of the double well structure, which stem from the majority species trapped in the minima of the lattice potential $V_A$. We find that for an interspecies interaction strength of $g_{AB}/E_rk_0^{-1}=1.0$ the density of the majority species distributes such that we obtain a nearly parity symmetric TAEP with respect to $x=0$ [cf. Figure \ref{fig:effective_potential} (d)].\par
In the following, we refer to the first, second, etc. maximum of the TAEP ordered from left to right excluding the case of $g_{AB}/E_rk_0^{-1}=0.2$ due to the small deviations of the TAEP from the double well structure.
Increasing the interspecies interaction strength eventually breaks the spatial symmetry of the TAEP w.r.t. $x=0$, which can be readily seen e.g. at $g_{AB}/E_rk_0^{-1}=2.0$ and $g_{AB}/E_rk_0^{-1}=4.0$ [cf. Figure \ref{fig:effective_potential} (e), (f)]. In particular, the TAEP for $g_{AB}/E_rk_0^{-1}=4.0$ exhibits a distinct asymmetry. Here, the second maximum of the TAEP is strongly suppressed compared to the other maxima. This can be attributed to the fact that the one-body density of the majority species $\rho_A^{(1)}$ in the second well of $V_A$, counted from the left, coincides with the position of the second maximum of the TAEP. Here, $\rho_A^{(1)}$ is strongly depopulated compared to the other wells of $V_A$ [cf. Figure \ref{fig:MB_densities} (h)] which leads to the suppression of the second maximum of the TAEP. Focusing on the maxima of the TAEP especially, on the third and fourth maximum, we can interpret these maxima as the potential barriers that the impurity has to overcome in order to tunnel from the left to the right side of the TAEP. We observe that the corresponding maxima heights increase with increasing $g_{AB}$, which indicates that the effective potential barrier for the impurity also increases with $g_{AB}$. \par
In the following, we investigate this effective potential barrier that separates the left from the right side of the TAEP. For this purpose, we calculate the relative difference $\Delta_{\textrm{max},i}=(\Lambda_i^\textrm{eff} - \Lambda_2^\textrm{eff})/\Lambda_2^\textrm{eff}$ between the second maximum height $\Lambda_2^\textrm{eff}$ and third and fourth maximum 
height, $\Lambda_3^\textrm{eff}$ and $\Lambda_4^\textrm{eff}$, of the TAEP, where $i=3,4$. Figure \ref{fig:effective_potential} (b) shows the relative difference $\Delta_{\textrm{max},i}$, which serves in the following as a measure for the effective potential barrier, in dependence on the interspecies interaction strength $g_{AB}$. For values below $g_{AB}/E_rk_0^{-1}=1.0$ the effective potential barrier $\Delta_{\textrm{max},i}\approx 0$ for $i=3,4$ meaning that the maxima barely deviate. For values above $g_{AB}/E_rk_0^{-1}=1.0$ the third and fourth maximum height of the TAEP become larger than the second one, breaking in this manner the symmetry of the TAEP. For large interspecies interaction strengths, e.g. $g_{AB}/E_rk_0^{-1}=4.0$, the effective potential barrier $\Delta_{\textrm{max},i}$ abruptly increases which is also due to the absence of $\rho_A^{(1)}$ in the second well of the lattice potential and, subsequently, the lowering of the second maximum height of the TAEP. The abrupt increase of the effective potential barrier $\Delta_{\textrm{max},i}$ intuitively leads to the assumption that for large interspecies interaction strengths, e.g. $g_{AB}/E_rk_0^{-1}=4.0$, a tunneling of the impurity from the left to the right TAEP should be strongly suppressed, as already seen in the one-body density [cf. Figure \ref{fig:MB_densities} (d)] obtained within the complete many-body approach.\par
Let us also describe the tunneling behavior of the impurity in terms of states that are highly localized in the minima of the TAEP. Since, for sufficiently strong interspecies interaction strengths $g_{AB}$ the TAEP resembles a lattice with five sites [cf. Figure \ref{fig:effective_potential} (a)] we calculate five of those functions. For this purpose, we construct an effective Hamiltonian $\hat{\mathcal{H}}_{\textrm{eff}}^{(B)} = 
- \frac{\hbar^2}{2m} \frac{\mathrm{d}^2}{(\mathrm{d}x^{B})^2}
+ \overline{V}_{\mathrm{eff}}^B(x^{B})$ using the TAEP. For the localized functions we use the notion of generalized Wannier functions \cite{kivelson1,kivelson2} which have the advantage that they can be also obtained in the presence of a non-periodic potential. To obtain five generalized Wannier functions $w_i^{(B)}$ we first calculate the five energetically lowest eigenfunctions of $\hat{\mathcal{H}}_{\textrm{eff}}^{(B)}$. Using these eigenfunctions as a basis we diagonalize the position operator $\hat{X}$ yielding eigenstates which are highly localized in the minima of the TAEP. For simplicity, we shall call the generalized Wannier functions in the following Wannier functions and, further, we will refer to the first, second, etc. Wannier function as the Wannier function localized in the first, second, etc. well of the TAEP.\par
Figure \ref{fig:effective_potential} (c)-(f) presents the absolute squares of the Wannier functions together with the TAEPs for the four different tunneling regimes corresponding to the interspecies interactions strengths $g_{AB}/E_rk_0^{-1}=0.2,1.0,2.0,4.0$. For $g_{AB}/E_rk_0^{-1}=0.2$ [Figure \ref{fig:effective_potential} (c)] we find, compared to the results for larger $g_{AB}$, the largest overlap between the Wannier functions. This indicates that those Wannier functions are rather ill-defined. The reason can be found in the TAEP which resembles for small interspecies interaction strengths, e.g. $g_{AB}/E_rk_0^{-1}=0.2$, more a double well than a lattice potential. Increasing $g_{AB}$ [cf. Figure \ref{fig:effective_potential} (d)-(f)], the Wannier functions become more localized in the minima of the TAEP and, therefore, are suited for the further analysis of the many-body wave function.\par
In summary, we have developed an effective one-body Hamiltonian using a TAEP for the purpose of constructing generalized Wannier functions from the eigenfunctions of this effective one-body Hamiltonian. This procedure resulted in functions which are highly localized in the wells of the TAEP for sufficiently large interspecies interaction strengths $g_{AB}$. Projecting these Wannier functions onto the full many-body wave function, obtained in the course of our numerical simulations, we are in the following able to get a deeper insight into the tunneling dynamics of the impurity.

\subsection{Dynamical response in terms of Wannier states}
In the following discussion, we analyze the results of the correlated many-body calculations utilizing the Wannier functions derived in the previous section more specifically. 
Therefore, we project the $i$-th Wannier function $w_i^{(B)}$ onto the many-body wave function and thereby receive the time-dependent  probability $P^{1,B}(w_i^{(B)})$ that the impurity occupies the $i$-th well of the TAEP. This probability is defined as
\begin{equation}
P^{1,B}(w_i^{(B)})=\abs{\braket{w_i^{(B)}}{\Psi_{\textrm{MB}}}}^2.
\end{equation}
Before discussing the results, let us comment on the Wannier functions as a basis set for the species wave function of the impurity. By summing up the five occupation probabilities of the Wannier functions, we obtain for each time instant of the evolution a measure for the accuracy of this basis representation. For our results we find that $\sum_i P^{1,B}(w_i^{(B)}) > 0.95$ so that we can consider the Wannier functions as an adequate basis set for describing the tunneling dynamics of the impurity.
\begin{figure*}[t]
	\centering
	\includegraphics[width=\textwidth]{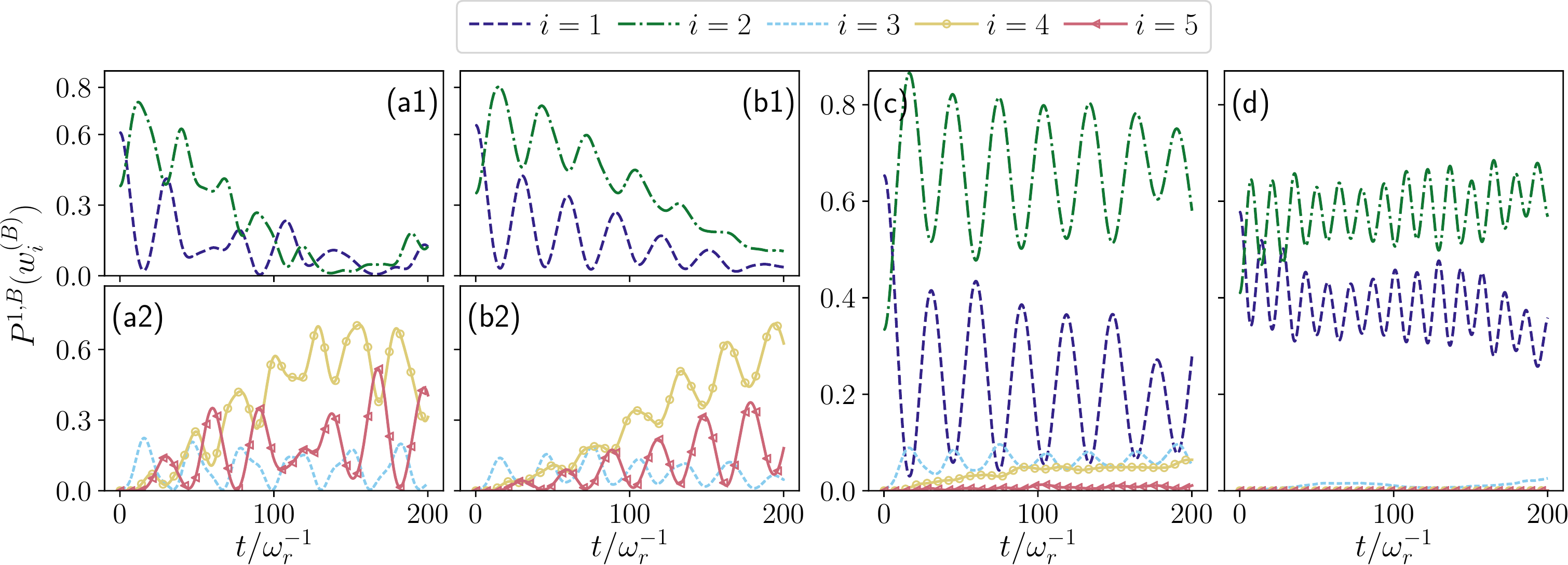}
	\caption{Probability $P^{1,B}(w_i^{(B)})$ of the many-body wave function to occupy a specific Wannier function (see legend), which is localized to one well in the corresponding TAEP. The probability $P^{1,B}(w_i^{(B)})$ is shown in panels for four different interspecies interaction strengths $g_{AB}/E_rk_0^{-1}=0.2,1.0,2.0,4.0$ in the panels (a)-(d), respectively. Panel (a1) and (a2) refer to $g_{AB}/E_rk_0^{-1}=0.2$, while panel (b1) and (b2) refer to $g_{AB}/E_rk_0^{-1}=1.0$. We consider a minority species consisting of $N_B=1$ particle and a majority species with $N_A=8$ particles which interact repulsively with $g_{AA}/E_rk_0^{-1}=1.0$.} 
	\label{fig:Wannier_states_occupation}
\end{figure*}
The time evolution of the probability $P^{1,B}(w_i^{(B)})$ for the four distinct tunneling regimes corresponding to $g_{AB}/E_rk_0^{-1}=0.2,1.0,2.0,4.0$ is shown in Figure \ref{fig:Wannier_states_occupation}. In the following, we aim at understanding the respective one-body densities $\rho_B^{(1)}$ obtained from the many-body calculation [see Figure \ref{fig:MB_densities} (a)-(d)] with the aid of the occupation probabilities $P^{1,B}(w_i^{(B)})$. Starting from $g_{AB}/E_rk_0^{-1}=0.2$ [Figure \ref{fig:Wannier_states_occupation} (a1), (a2)], we observe a rather irregular time evolution of the occupation probabilities. Here, many Wannier functions are occupied simultaneously indicating that the state of the impurity consists of a superposition of the corresponding Wannier functions. At first glance the behavior of the time evolution of $P^{1,B}(w_i^{(B)})$ at $g_{AB}/E_rk_0^{-1}=1.0$ depicted in Figure \ref{fig:Wannier_states_occupation} (b1), (b2) is similar compared to the one in panel (a1), (a2). In both cases we observe a reduction of the occupation probabilities of the first and second Wannier function, representing the left side of the TAEP, and an increase of $P^{1,B}(w_i^{(B)})$ of the fourth and fifth Wannier function, representing the right side of TAEP. This gradual depopulation of the left and subsequent population of the right side of the TAEP reflects precisely the observed tunneling process observed in the one-body density $\rho_B^{(1)}$ of the impurity. However, we find that at $g_{AB}/E_rk_0^{-1}=1.0$ the transfer of the occupation probabilities from the left to the right side of the TAEP is more uniform than at $g_{AB}/E_rk_0^{-1}=0.2$. \par
Initially, we observe an exchange of probability between the first and second Wannier states [cf. Figure \ref{fig:Wannier_states_occupation} (b1)], which represents the material barrier tunneling process in the initial site of the double well. Eventually, probability is transferred from the first and second to the fourth and fifth Wannier state [cf. Figure \ref{fig:Wannier_states_occupation} (b2)], reflecting the controlled tunneling behavior observed in Figure \ref{fig:MB_densities} (b). Partially, this can be understood in terms of the TAEP which exhibits a lattice structure on top of the double well, while being still spatially symmetric with respect to $x=0$.
For stronger interspecies interaction strengths, e.g. $g_{AB}/E_rk_0^{-1}=2.0$ and $g_{AB}/E_rk_0^{-1}=4.0$, the first two Wannier functions are predominantly occupied. As shown in Figure \ref{fig:Wannier_states_occupation} (c), for times directly after the quench the first Wannier function is the most occupied one, whereas for larger times the occupation probability for the second Wannier function becomes the dominant one. We can understand this intuitively by inspecting the corresponding TAEP depicted in Figure \ref{fig:effective_potential} (e). Here, the TAEP exhibits a global minimum in the second well which makes it energetically favorable for the impurity to reside there. Furthermore, the first and second occupation probability exhibit a strong counterwise oscillation behavior. Comparing this with the oscillation of the one-body density of the impurity around a potential barrier imposed by the majority species [see Figure \ref{fig:MB_densities} (c)], we find very good agreement. Moreover, we find that the occupation probability for the other Wannier functions, i.e. the third fourth and fifth, are strongly suppressed, but seem to continuously increase over time. Therefore one might assume that for sufficiently long evolution times the impurity eventually tunnels to the right side of the TAEP.\par
Turning to the occupation probabilities at $g_{AB}/E_rk_0^{-1}=4.0$ [Figure \ref{fig:Wannier_states_occupation} (d)], we observe that the occupation probabilities of the third, fourth and fifth Wannier function are close to zero during the entire time propagation. We can understand this in terms of an effective potential barrier that separates the left side of the TAEP from the right side w.r.t. $x=0$. Here, the value for the relative difference $\Delta_{\textrm{max},i}$ for $g_{AB}/E_rk_0^{-1}=4.0$ is much larger than for the other considered interspecies interaction strength $g_{AB}$ [see Figure \ref{eq:effective_potential} (b)] and, therefore, it is very unlikely that the impurity will tunnel to the right side of the TAEP even for later times. Additionally, the occupation probabilities of the first and second Wannier function perform a counterwise oscillation, likewise to the probabilities at $g_{AB}/E_rk_0^{-1}=2.0$ [Figure \ref{fig:Wannier_states_occupation} (c)]. In contrast to the aforementioned oscillation of the occupation probabilities, we observe for $g_{AB}/E_rk_0^{-1}=4.0$ almost twice the number of oscillation periods during the dynamics as well as a reduction of the amplitudes. A hint for understanding this gives again the corresponding TAEP [see Figure \ref{eq:effective_potential} (f)]. Here, the second maximum height is strongly suppressed and the first and second  Wannier functions have a large overlap. From this we can infer that the species wave function of the impurity consists mainly of a superposition of the first and second Wannier function whose contributions oscillate over time.\par
In conclusion, we generated localized Wannier functions associated with the TAEP and projected them onto the time-dependent many-body wave function. Having this at hand we were able to describe the many-body impurity dynamics in terms of the evolution of the occupation probabilities of the respective Wannier functions.
\begin{figure*}[t]
	\centering
	\includegraphics[width=0.9\textwidth]{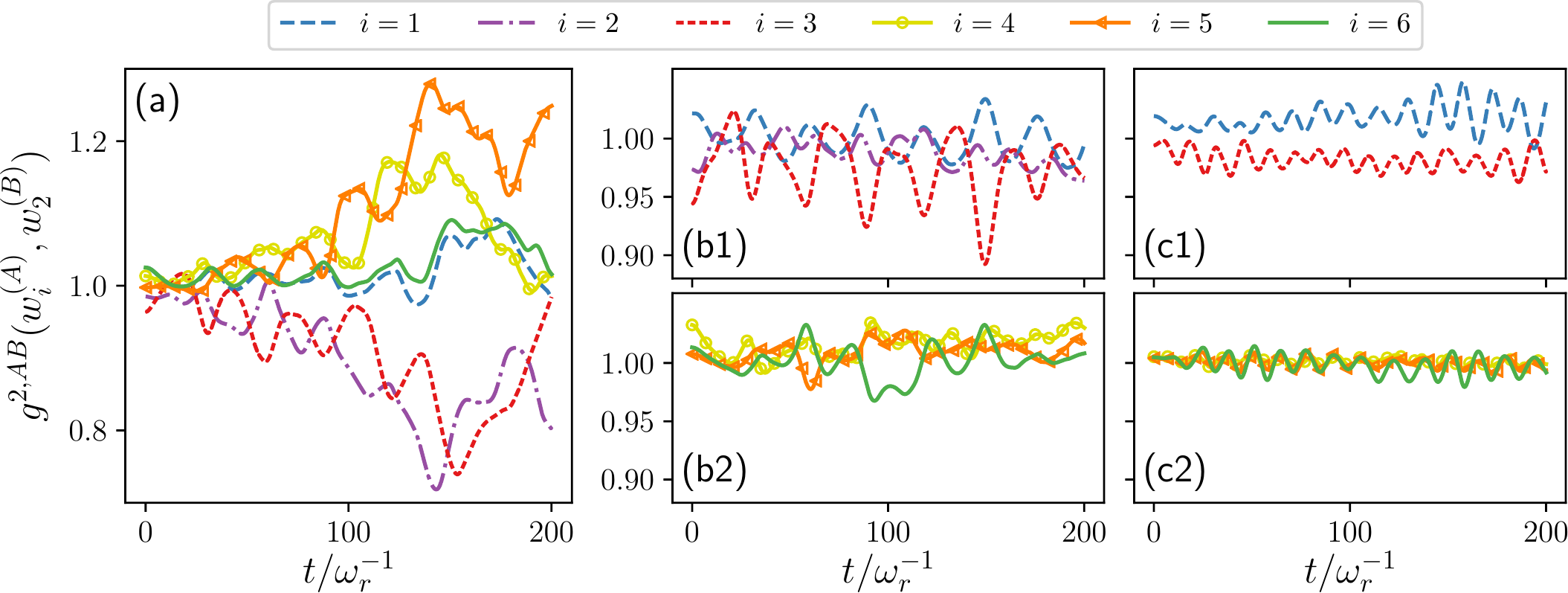}
	\caption{Temporal evolution the discrete two-body correlations $g^{2,AB}(w_i^{(A)},w_2^{(B)})$ between the impurity occupying the second site of the TAEP and one particle of the majority species occupying the $i$-th site of the lattice for interspecies interaction strengths (a) $g_{AB}/E_rk_0^{-1}=1.0$, (b1), (b2) $g_{AB}/E_rk_0^{-1}=2.0$ and (c1), (c2) $g_{AB}/E_rk_0^{-1}=4.0$. In panel (c1) we explicitly exclude the results for $g^{2,AB}(w_2^{(A)},w_2^{(B)})$, because here the occupation probability $P^{1,A}(w_2^{(A)})$ is very small during the time propagation leading to an artificial amplification of $g^{2,AB}(w_2^{(A)},w_2^{(B)})$ [see Eq. \ref{eq:g2}]. We consider a minority species consisting of $N_B=1$ particle and a majority species consisting of $N_A=8$ particles which interact repulsively with $g_{AA}/E_rk_0^{-1}=1.0$.}
	\label{fig:discrete_correlation_fct}
\end{figure*}
As a natural next step, we shall investigate the two-body correlations between the impurity and the majority species. More precisely, we determine the discrete correlation function associated with the impurity occupying a specific well of the TAEP and a single particle of the majority species occupying a certain well of the lattice potential $V_A$.\par
First we generate Wannier functions associated with the single-particle Hamiltonian for the majority species
$\hat{\mathcal{H}}^{(A)} =
- \frac{\hbar^2}{2m} \frac{\mathrm{d}^2}{(\mathrm{d}x^{A})^2}
+ V_A(x^A)$,
following the procedure explained above \footnote{Note that the Wannier functions are sorted from the left to right regarding the sites of the lattice potential.}. Furthermore, we determine the one-body probability for a particle of the majority species to occupy the $j$-th well of the lattice potential $V_A$ by projecting the many-body wave function onto the Wannier function $w_j^{(A)}$, thereby constructing the probability $P^{1,A}(w_i^{(A)})=\abs{\braket{w_i^{(A)}}{\Psi_{\textrm{MB}}}}^2$.
The conditional probability $P^{2,AB}(w_i^{(A)},w_j^{(B)}) =\langle\Psi_{\text{MB}}| \hat{O}^{(2)}_{ij}|\Psi_{\text{MB}}\rangle$ of finding a single particle of the majority species in the $i$-th well of $V_A$ and at the same time the impurity in the $j$-th well of the TAEP is defined as the expectation value of the following operator 
\begin{equation}
\hat{O}^{(2)}_{ij}=\frac{1}{N_A}\sum^{N_A}_{l} |w^{(A),l}_i\rangle\langle w^{(A),l}_i| \otimes  |w^{(B)}_j\rangle\langle w^{(B)}_j|,
\end{equation}
with respect to the many-body wave function $|\Psi_{\text{MB}}\rangle$. The summation runs over the number of particles of the subsystem A.
The discrete two-body correlation function is then given by
\begin{equation}
g^{2,AB}(w_i^{(A)},w_j^{(B)}) = \frac{P^{2,AB}(w_i^{(A)},w_j^{(B)})}{ P^{1,A}(w_i^{(A)}) P^{1,B}(w_j^{(B)})}.
\label{eq:g2}
\end{equation}
This function provides information about the correlation between a single particle of the majority species localized at the $i$-th well of $V_A$ and the impurity species localized at the $j$-th well of the TAEP. In case the discrete correlation function $g^{2,AB}(w_i^{(A)},w_j^{(B)})$ equals unity the particle of the majority species and the impurity are termed uncorrelated since the conditional probability equals the unconditional one. However, if $g^{2,AB}(w_i^{(A)},w_j^{(B)})$ is larger (smaller) than unity the impurity and the particle of the majority species are said to be correlated (anti-correlated) \cite{simos1,mlb2}.\par
The time evolution of the discrete correlation function for the impurity occupying the second Wannier function $w_2^{(B)}$ of the TAEP and the particle of the majority species occupying one of the six Wannier functions of the lattice potential $V_A$ are shown in Figure \ref{fig:discrete_correlation_fct}. We choose to present only results for $g^{2,AB}$ where the impurity is occupying the second Wannier state $w_2^{(B)}$ since for this case we are already able to observe and analyze all relevant properties of the discrete correlation function. Figures \ref{fig:discrete_correlation_fct} (a)-(c2) depict $g^{2,AB}$ for the interspecies interaction strengths $g_{AB}/E_rk_0^{-1}=1.0,2.0,4.0$, respectively \footnote{Here, we do not show the results for $g_{AB}/E_rk_0^{-1}=0.2$, because as already mentioned before the Wannier basis is ill-defined in this case.}.
%Figure 8a
At $g_{AB}/E_rk_0^{-1}=1.0$ [Figure \ref{fig:discrete_correlation_fct} (a)], we observe that initially the system is rather uncorrelated and develops stronger correlations for larger time. Here, the particle of the majority species in the fourth/fifth well of $V_A$ and the impurity species in the second well of $\overline{V}^B_{\mathrm{eff}}$ show an increasing correlation amplitude over time, whereas an anti-correlation between the particle of the majority species in the second/third well of $V_A$ and the impurity species in the second well of $\overline{V}^B_{\mathrm{eff}}$ occurs. The particle of the majority species being in the first and sixth well exhibits a similar correlation behavior with the impurity in the second well. We note that the strongest correlation (anti-correlation) occurs within the time intervals where the impurity tunnels to the right side of the TAEP [cf. Figure \ref{fig:MB_densities} (b)].\par
Turning to the discrete correlation functions at $g_{AB}/E_rk_0^{-1}=2.0,4.0$, depicted in Figure \ref{fig:discrete_correlation_fct} (b1), (b2) and (c1), (c2)  we find that the system is in both cases slightly two-body correlated. Predominantly, this is the case for $g_{AB}/E_rk_0^{-1}=4.0$ where the impurity exhibits a self-trapping behavior, showing only a weak distinct correlated or anti-correlated behavior between the impurity in the second well of the TAEP and one particle of the majority species in a specific well of the lattice potential [cf. Figure \ref{fig:discrete_correlation_fct} (c1), (c2)]. However, in Figure \ref{fig:discrete_correlation_fct} (b1) we observe an oscillating correlation behavior between the majority species in the third well of $V_A$ and the impurity species in the second well of $\overline{V}^B_{\mathrm{eff}}$. Here, $g^{2,AB}(w_3^{(A)},w_2^{(B)})$ oscillates between the anti-correlated and uncorrelated case which might be associated with the material barrier tunneling of the impurity in the initial site of the double well [cf. Figure \ref{fig:MB_densities} (c)].
Concluding, we observe an overall decrease of the discrete correlation with increasing interspecies interaction strength, which appears to be related to the manifestation of the self-trapping of the impurity.

%% file: two_impurities.tex
% !TeX root = ../paper_dwcl.tex
\section{Correlated tunneling dynamics of two impurities}
\label{sec:two_impurities}
\begin{figure*}[t]
	\centering
	\includegraphics[scale=0.69]{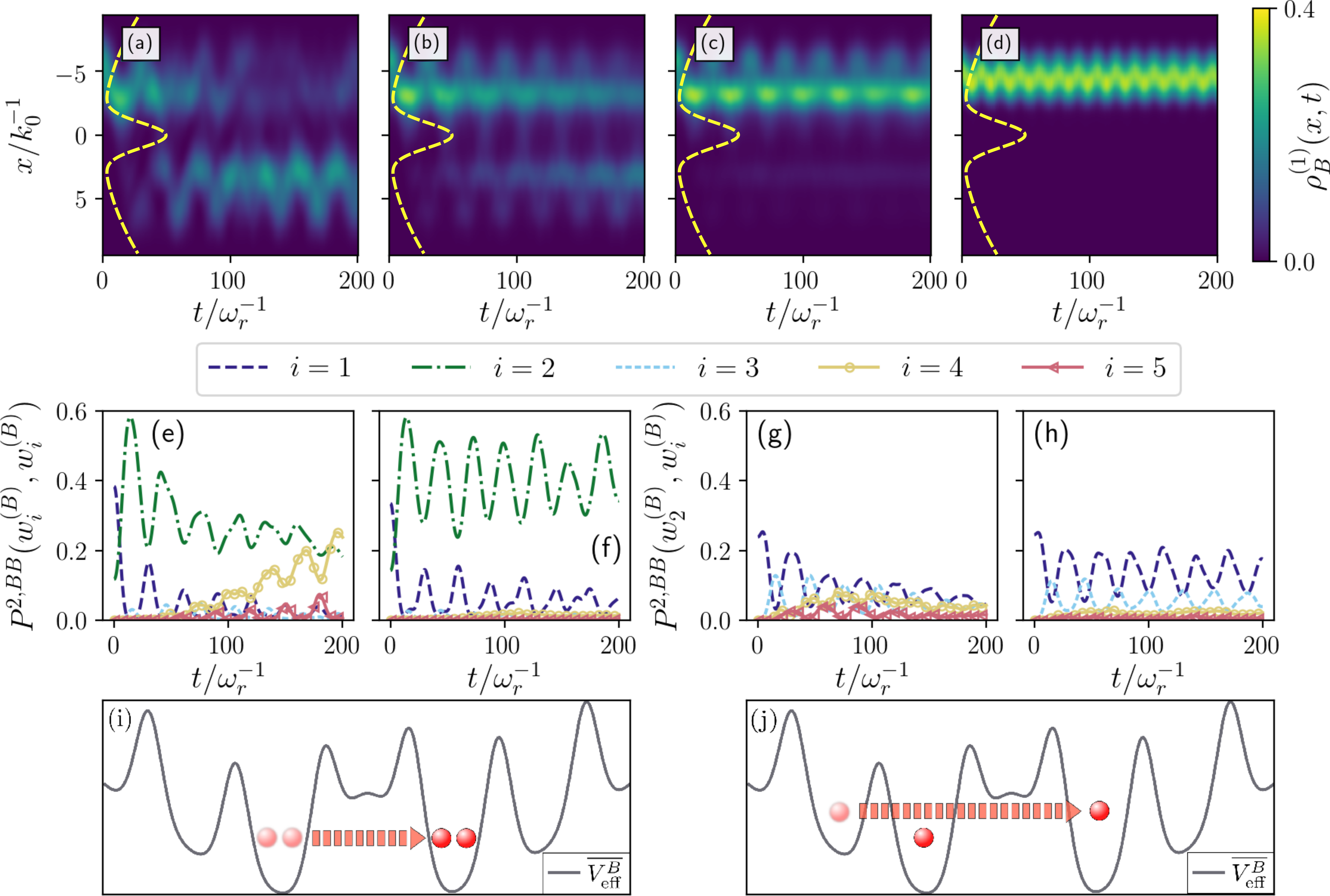}
	\caption{(a)-(d) Temporal evolution of the one-body density of the two impurities for $g_{AB}/E_rk_0^{-1}=0.2,0.8,1.3,4.0$, respectively. The dashed yellow line represents the double well potential for the impurity species. Conditional probability $P^{2,BB}(w_i^{(B)}, w_j^{(B)})$ to find one impurity in the Wannier state $w_i^{(B)}$ associated with the TAEP and another impurity in $w_j^{(B)}$ at (e), (g) $g_{AB}/E_rk_0^{-1}=0.8$ and at (f), (h) $g_{AB}/E_rk_0^{-1}=1.3$. (i) Sketch of a pair-particle tunneling process and (j) of a single-particle tunneling process w.r.t. the TAEP for $g_{AB}/E_rk_0^{-1}=0.8$.}
	\label{fig:NB2_multi}
\end{figure*}
So far we have investigated the tunneling dynamics of a single impurity coupled to a majority species and found that the tunneling behavior can be steered by varying the interspecies interaction strength $g_{AB}$.
In the following we unravel whether a similar controlled tunneling process can be realized employing two impurities. For this purpose, we consider two weakly interacting impurities, i.e. they interact repulsively via a contact potential of strength $g_{BB}/E_rk_0^{-1}=0.2$, embedded in the same potential setup as shown in Figure \ref{fig:Setup}. The parameters for the majority species remain unchanged compared to the single impurity case, i.e. $N_A=8$ and $g_{AA}/E_rk_0^{-1}=1.0$. The tunneling process is induced by performing the same quench protocol as in the case of a single impurity. Due to a tilting potential $V_\textrm{tilt}$ the minority species is initially prepared in the left side of the double well $V_B$. Setting $V_\textrm{tilt}$ suddenly to zero we monitor the respective tunneling dynamics.\par
Figure \ref{fig:NB2_multi} (a)-(d) presents the one-body densities $\rho_B^{(1)}(x,t)$ of the two impurities for varying interspecies interaction strengths. As it can be seen, the dynamics of $\rho_B^{(1)}(x,t)$ for $N_B=2$ resemble the one-body densities in the case of a single impurity [cf. Figure \ref{fig:MB_densities} (a)-(d)]. For weak interspecies interaction strengths, i.e. $g_{AB}/E_rk_0^{-1}=0.2$, we again observe a rather irregular tunneling dynamics of the impurity species [cf. Figure \ref{fig:NB2_multi} (a)]. Increasing $g_{AB}$, also for $N_B=2$ the impurity species performs a material barrier tunneling within the initially populated well and finally tunnels to the other well [cf. Figure \ref{fig:NB2_multi} (b)]. A further increase of $g_{AB}$ finally yields a self-trapping behaviour of the impurity species due to the strong repulsion [cf. Figure \ref{fig:NB2_multi} (d)]. As a result also a system with two impurities exhibits the aforementioned four tunneling regimes. However, introducing an additional impurity to the system leads to a shift of the tunneling regimes [cf. Figure \ref{fig:NB2_multi} (b), (c)] with respect to the interspecies interaction strength ($g_{AB}/E_rk_0^{-1}=0.2,0.8,1.3,4.0$). Therefore, we can conclude that also for two impurities we are able to control the quench induced tunneling process via the coupling to the majority species.\par
Having identified the existence of the four tunneling regimes of the impurity species the question that arises is whether the impurities tunnel pairwise through the potential landscape, which we would refer to as pair tunneling, or whether they tunnel individually, which we would call single particle tunneling. 
To expose the underlying mechanisms we present in Figure \ref{fig:NB2_multi} (e)-(h) the conditional probability $P^{2,BB}(w_i^{(B)}, w_j^{(B)})$ of detecting one impurity in the Wannier state $w_i^{(B)}$ and the other impurity in the Wannier state $w_j^{(B)}$. Here, we will refer to $w_i^{(B)}$ as the generalized Wannier states associated with the TAEP which we introduced in section \ref*{sec:Effective potential} [see also Eq. (\ref{eq:TAEP})].
Figure \ref{fig:NB2_multi} (e)-(f) shows the conditional probability $P^{2,BB}(w_i^{(B)}, w_i^{(B)})$ that the two impurities occupy the same Wannier state for interspecies interaction strengths $g_{AB}/E_rk_0^{-1}=0.8$ [Figure \ref{fig:NB2_multi} (e)] and $g_{AB}/E_rk_0^{-1}=1.3$ [Figure \ref{fig:NB2_multi} (f)]. Thus, we focus on the two cases where a material barrier tunneling of the impurity species takes place or where the latter process is accompanied by a subsequent tunneling to the other site of the double well.
In Figure \ref{fig:NB2_multi} (e) we observe a decreasing probability $P^{2,BB}(w_i^{(B)}, w_i^{(B)})$ to find both impurities in the second Wannier state, whereas the probability to detect both impurities in the fourth Wannier state increases. This probability transfer indicates a pair tunneling from the left to the right side of the TAEP w.r.t. $x=0$. Additionally, the impurities perform the material barrier tunneling as a pair, which can be inferred from the alternating increase and decrease of $P^{2,BB}(w_1^{(B)}, w_1^{(B)})$ and $P^{2,BB}(w_2^{(B)}, w_2^{(B)})$.
A schematic representation of this process is illustrated in Figure \ref{fig:NB2_multi} (i) assuming the TAEP at $g_{AB}/E_rk_0^{-1}=0.8$.
For the investigation of the single particle tunneling we show in Figure \ref{fig:NB2_multi} (g), (h) the conditional probability to find one impurity in the second Wannier state and the other impurity in another Wannier state for the above-mentioned $g_{AB}$.
A decrease of the probability to find one impurity in the second and one impurity in the first Wannier state $P^{2,BB}(w_2^{(B)}, w_1^{(B)})$ is observed for times up to $t/\omega_r^{-1}=100$ [Figure \ref{fig:NB2_multi} (g)], while the conditional probability $P^{2,BB}(w_2^{(B)}, w_4^{(B)})$ increases. This suggests a tunneling of one impurity from the first well of the TAEP to the fourth well, whereas the other impurity remains in the second well. This process is depicted in Figure \ref{fig:NB2_multi} (j). We note that this is one of many single particle tunneling processes that can take place. In this sense, the tunneling process of the impurity species is rather complex, consisting of single particle and pair tunneling processes.\par
Concluding we have realized the four tunneling regimes which we previously identified in Figure \ref{fig:MB_densities} (a)-(d) also for two weakly interacting impurities coupled to a majority species.
This implies that it is also possible to control the tunneling process of two impurities via the interspecies interaction strength. Eventually, we have characterized the tunneling processes underlying the dynamical response of the impurity species in terms of single particle and pair tunneling processes \cite{erdmann}.

%% file: conclusion.tex
\section{Conclusions and Outlook}
\label{sec:Conclusion}

We have investigated the correlated tunneling dynamics of impurities trapped in a double well potential and immersed in a lattice trapped majority species. The tunneling dynamics was initiated by implementing an initial tilt of the double well, thereby localizing the impurity species in one of the wells, and quenching this to a symmetric potential configuration. In case of a single impurity we have identified four different tunneling regimes w.r.t. the interspecies interaction strength. For very weak interspecies interaction strengths the tunneling of the impurity can be characterized as rather complex, exhibiting no regular or repetitive structure. However, increasing the coupling to the majority species leads to a regular tunneling behavior of the impurity, which consists of an initial material barrier tunneling due to the presence of the majority species and is followed by a transfer of the impurity to the other site of the double well. Additionally, this effect is accompanied by a strong entanglement between the subsystems. A further increase of the interspecies interaction strength leads to a sole material barrier tunneling in the initial site of the double well for long time intervals and finally for very large couplings forces the impurity to localize in the initially populated well and being self-trapped.\par
In order to gain insight into the underlying microscopic processes of the emergent correlated tunneling dynamics, we have constructed a time-averaged effective potential (TAEP) based on the one-body density of the majority species. Depending on the interspecies interaction strength, this effective potential exhibits an additional structure in each site of the double well, thus explaining the material barrier tunneling. Increasing the coupling to the majority species, the TAEP is predominantly formed by the one-body density of the majority species and the presence of the double well is of minor consequence, resulting in the observed self-trapping of the impurity. Moreover, the generalized Wannier states associated with this potential allowed for a characterization of the impurity's dynamical response as well as the involved correlations. We concluded our study with an investigation of two weakly repulsively interacting impurities which we prepared analogously to the case of a single impurity. We were able to identify the previous four tunneling regimes for smaller interspecies interaction strengths, being shifted to $g_{AB}/E_rk_0^{-1}=0.2,0.8,1.3,4.0$ respectively, compared to the scenario of a single impurity. Employing again the TAEP we have developed an understanding of the tunneling dynamics, which consists of a superposition of pair tunneling as well as single particle tunneling processes.\par
There are various interesting research directions that prove to be promising for future investigations relying on the findings of the current work. A direct extension involves the inclusion of spin degrees of freedom between the impurities. Here, the possible formation of an analogue of a Cooper-pair in the course of the tunneling dynamics would be of immediate interest. Another straightforward direction would be to consider quench protocols which also include a variation of the interspecies interaction strengths. For example, one might think of a subsequent interaction quench after a transfer of the impurity in order to prevent tunneling to the initially populated site. Also, dynamically driving the corresponding parameters of the system might be useful for transferring the impurity species in a more controlled and systematic manner.\par

%% file: appendix.tex
\appendix
\section{Tunneling dynamics for smaller tilting strength}
\begin{figure*}[t!]
	\centering
	\includegraphics[scale=0.55]{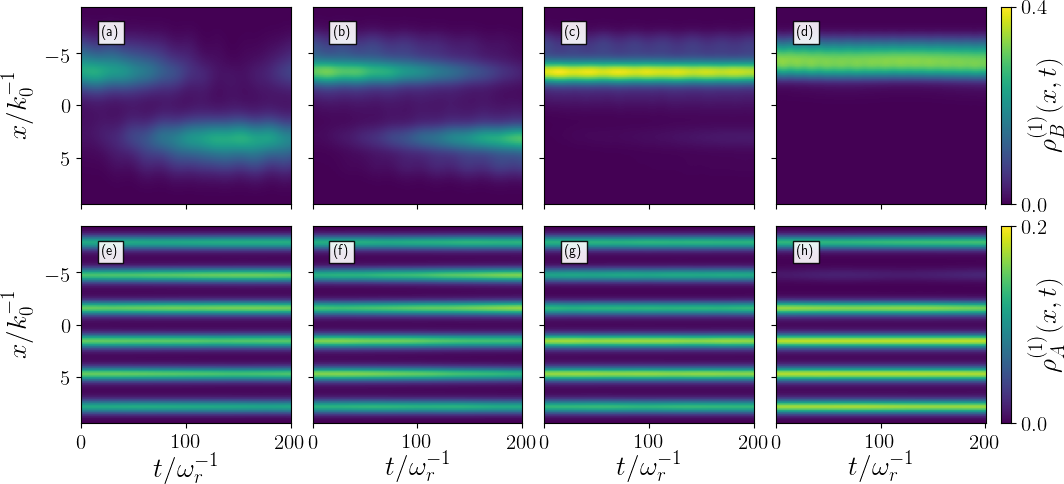}
	\caption{One-body density of (a)-(d) the impurity and (e)-(h) the majority species using a tilting strength $\alpha/E_Rk_0^{-1}=0.01$ within the full many-body approach. The results in each column correspond to the same interspecies interaction strength $g_{AB}$, ordered from left to right with $g_{AB}/E_rk_0^{-1}=0.2,1.0,2.0,4.0$. We considered for the majority species $N_A=8$ particles which interact repulsively with an intraspecies interaction strength of $g_{AA}/E_rk_0^{-1}=1.0$}
	\label{fig:MB_densities_small_tilt}
\end{figure*}
In the following we demonstrate that a certain minimal tilting strength $\alpha$ is necessary for observing the tunneling dynamics as in Figure \ref{fig:MB_densities} where $\alpha/E_Rk_0^{-1}=0.1$.
Figure \ref{fig:MB_densities_small_tilt} shows the temporal evolution of the one-body densities of the impurity [Figure \ref{fig:MB_densities_small_tilt} (a)-(d)] and the majority species [Figure \ref{fig:MB_densities_small_tilt} (e)-(h)] using a tilting strength $\alpha/E_Rk_0^{-1}=0.01$, within the full many-body approach. Analogously to the previous discussion in section \ref{sec:Tunneling Dynamics}, we induce the dynamics by initially tilting the double well $V_B$ of the impurity with a tilting strength $\alpha$ and let the system evolve in time for $\alpha=0$. However, in the present case lowering the initial tilting strength to $\alpha/E_Rk_0^{-1}=0.01$ leads to a smaller initial energy offset between the sites of $V_B$.\par
For weak $g_{AB}$, i.e. $g_{AB}/E_rk_0^{-1}=0.2,1.0$, we find a rather regular tunneling of the impurity from the left to the right side of the double $V_B$ [cf. Figure \ref{fig:MB_densities_small_tilt} (a), (b)]. Comparing this with the dynamical response of an impurity for an initial tilting strength of $\alpha/E_Rk_0^{-1}=0.1$ [cf. Figure \ref{fig:MB_densities} (b)] we find no material barrier tunneling triggered by the density of the majority species. The difference between the two initial tilting strengths is also evident for larger interspecies interaction strengths, e.g. $g_{AB}/E_rk_0^{-1}=2.0,4.0$. Here, the impurity essentially remains localized throughout the dynamics and does not perform any oscillations [cf. Figure \ref{fig:MB_densities_small_tilt} (c), (d)]. Furthermore, the one-body density of the majority species behaves accordingly and does not exhibit a distinctive dynamics compared to the $\alpha/E_Rk_0^{-1}=0.1$. Namely, $\rho_A^{(1)}$ remains well localized at the sites of the lattice potential during the propagation [cf. Figure \ref{fig:MB_densities_small_tilt} (e)-(h)].
These observations lead to the conclusion that indeed a sufficiently high initial tilting strength $\alpha$ is needed in order to observe a material barrier tunneling with a subsequent controlled transfer of the impurity to the other side of the double well.

\section{Dependence of the tunneling process on the system parameters}
\label{appB}
\begin{figure*}[t]
	\includegraphics[width=\textwidth]{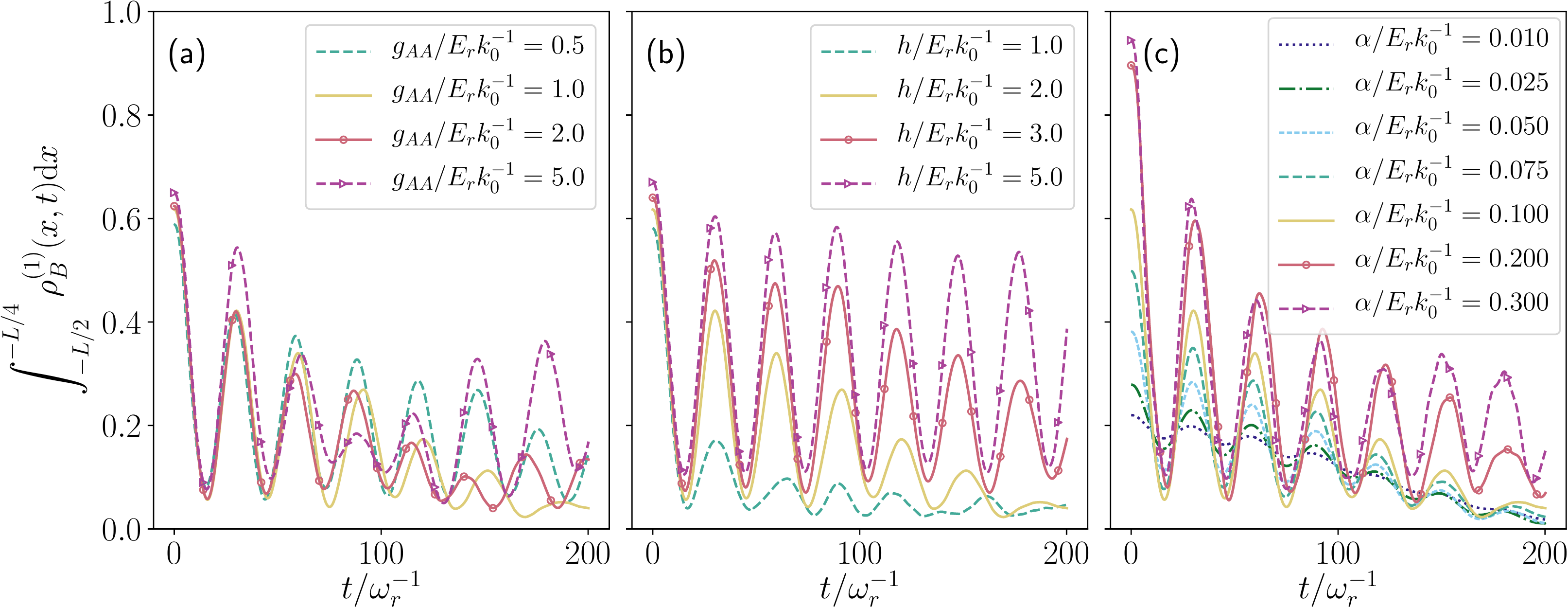}
	\caption{Temporal evolution of the integrated one-body density of the impurity $\int_{-L/2}^{-L/4}\rho_B^{(1)}(x,t) dx$ upon variation of (a) the intraspecies interaction strength $g_{AA}$, (b) the barrier height $h$ and (c) the tilt $\alpha$ for an interspecies interaction strength of $g_{AB}/E_R k^{-1}_0=1.0$. For each variation the remaining parameters have been fixed to the values as introduced in section {\ref{sec:setup_methodology}} A. The yellow lines correspond to the parameter choice in the main text. We consider a minority species consisting of $N_B=1$ particle and a majority species of $N_A=8$ particles which interact repulsively with $g_{AA}/E_rk_0^{-1}=1.0$.}
	\label{fig:parameter_scan}
\end{figure*}
We have analyzed the tunneling behavior of the impurity species for a specific choice of the intraspecies interaction strength $g_{AA}$ of the majority species and the barrier height $h$. In Figure {\ref{fig:parameter_scan}} we show that the qualitative behavior of the tunneling dynamics discussed in the main text can be recovered for significantly varying $g_{AA}$ and $h$. As a measure for the characteristic dynamical response of the impurity we again investigate the temporal evolution of the integrated one-body density of the impurity $\int_{-L/2}^{-L/4}\rho_B^{(1)}(x,t) dx$. The latter enables us to distinguish between the different tunneling regimes, for a fixed interspecies interaction strength of $g_{AB}/E_R k^{-1}_0=1.0$. This value lies in regime II in Figure {\ref{fig:integrated_density}} (a), where we observe a material barrier tunneling within the initially populated well with a final transport of the impurity to the other site of the double well. In Figure {\ref{fig:parameter_scan}} (a) we observe that an increase of $g_{AA}$ leads to a faster revival of the material barrier tunneling in the initially populated well. Decreasing $g_{AA}$ leads to a temporal prolongation of the material barrier tunneling and thereby a delayed transfer of the impurity to the other site of the double well. \par
A similar process can be observed when increasing the height of the double well barrier. However, at a certain height of the barrier, e.g. $h/E_rk_0^{-1}=5.0$, the impurity is barely able to tunnel to the other site of the double well within the considered time interval and solely performs the material barrier tunneling in the initial well. In contrast, a sufficiently small barrier height, e.g. $h/E_rk_0^{-1}=1.0$, leads to a less dominant material barrier tunneling within the initial well since the impurity can be easily transferred to the other well. \par
Finally, we have investigated the dependence of the impurity's dynamical response on the initial tilt $\alpha$ [cf. Figure {\ref{fig:parameter_scan}} (c)]. For small tilts we find almost no oscillations in the initially populated well and the impurity is directly transferred to the other site of the double well [see also Figure {\ref{fig:MB_densities_small_tilt}} (b)]. For increasing tilts $\alpha$ the oscillations due to the material barrier tunneling become more prominent and thereby the tunneling to the other well is delayed.
In this sense, we find that the material barrier tunneling can be recovered for various parameters of the system.

\section{Convergence of the many-body simulations}
\begin{figure}[t]
	\includegraphics[width=.45\textwidth]{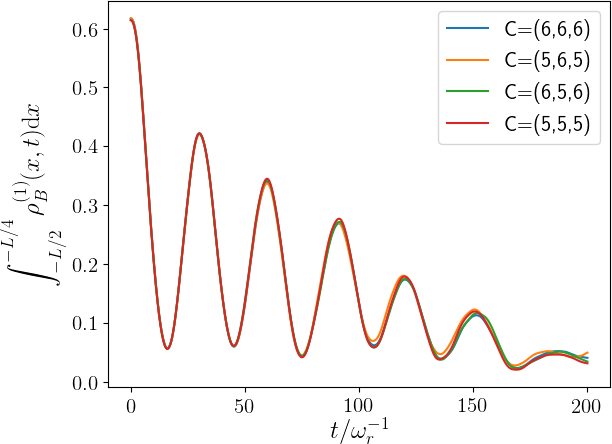}
	\caption{Temporal evolution of the integrated one-body density of the impurity $\int_{-L/2}^{-L/4}\rho_B^{(1)}(x,t) dx$ upon varying the orbital configuration $C$ for an interspecies interaction strength of $g_{AB}/E_R k^{-1}_0=1.0$. The system consists of a minority species with $N_B=1$ particle and a majority species of $N_A=8$ particles which interact repulsively with $g_{AA}/E_rk_0^{-1}=1.0$.}
	\label{fig:convergence}
\end{figure}
In the following we briefly discuss the convergence behavior of our results. As discussed in section {\ref{sec:setup_methodology}} the size of the Hilbert space is given in terms of the orbital configuration $C=(M,d_A,d_B)$. Here, $M$ describes the number of species functions in the Schmidt representation (see Eq. {\ref{eq:schmidt}}), while $d_\sigma$ with $\sigma \in \{A,B\}$ refer to the number of single-particle functions building the time dependent number states $|\vec{n}^\sigma(t)\rangle$ (see Eq. {\ref{eq:ns}}). In the process of increasing the number of species functions and single-particle functions it is possible to recover the solution of the many-body wave function with an increasing accuracy. Due to the exponentially increasing size of the Hilbert space it is computationally prohibitive to use too many species and single-particle functions. However, we are able to obtain numerical solutions which incorporate all the necessary correlations and go beyond mean-field approximations utilizing ML-MCTDHX. We determine the effect of the truncation of the Hilbert space by investigating as a representative example the integrated one-body density of the impurity $\int_{-L/2}^{-L/4}\rho_B^{(1)}(x,t) dx$ upon varying the orbital configuration $C$. In Figure {\ref{fig:convergence}} we show the latter for an interspecies interaction strength of $g_{AB}/E_R k^{-1}_0=1.0$ [cf. Figure {\ref{fig:MB_densities}} (b)]. Note that $g_{AB}/E_R k^{-1}_0=1.0$ lies in the interval where the degree of correlations is maximized (cf. Figure {\ref{fig:depletion}}). As it can be seen, increasing the size of the Hilbert space systematically it is possible to achieve convergence. Based on these findings the orbital configuration $C=(6,6,6)$ has been employed in all many-body calculations presented in the main text.

%% file: acknowledgements.tex
\acknowledgements
P. S. gratefully acknowledges funding by the Deutsche Forschungsgemeinschaft in the framework of the SFB 925 "Light induced dynamics and control of correlated quantum systems". K. K. gratefully acknowledges a scholarship of the Studienstiftung des deutschen Volkes. S. I. M gratefully acknowledges financial support in the framework of the Lenz-Ising Award of the University of Hamburg. \par
\vspace{0.5cm}
F. T. and K. K. contributed equally to this work.